\documentclass[aip,graphicx]{revtex4-1}
\usepackage{graphics,amsmath,amssymb,graphicx}
\usepackage{xcolor}
\draft 

\begin{document}


\title{Not-so-efficient proton acceleration by an intense laser pulse obliquely incident on a structured target} 



\author{Imran Khan}
\email[]{imran.khan@physics.iitd.ac.in}
\affiliation{Department of Physics, Indian Institute of  Technology Delhi, Hauz Khas, New Delhi, India-110016}

\author{Vikrant Saxena}
\email[]{vsaxena@physics.iitd.ac.in}
\affiliation{Department of Physics, Indian Institute of  Technology Delhi, Hauz Khas, New Delhi, India-110016}

\date{\today}

\begin{abstract}
The interaction of an obliquely incident laser pulse with a flat TNSA target is known to enhance the cut-off energy of protons/ions as compared to a normally incident laser pulse, mostly owing to Brunnel absorption. However, it is not well understood how the oblique incidence of the laser pulse would affect the protons/ion spectra in the case of a micro-structured target. Using two-dimensional particle-in-cell simulations, we show here that the protons/ions cut-off energies are rather reduced in the case of oblique incidence of the laser pulse if the target has a micron-sized groove on its front surface. This is also found to be true for a periodically grooved target.

\end{abstract}

\maketitle 
\section{Introduction}

The energetic ions/protons generated through the interaction of a highly intense femtosecond laser pulse with a solid target have recently attracted wide interest due to their potential applications, such as in isochoric heating \cite{patel2003isochoric}, radiographic density diagnostic \cite{mackinnon2006proton}, hadron therapy \cite{bulanov2002feasibility, ledingham2014towards}, fast ignition \cite{roth2001fast, atzeni2002first}, probing short-lived magnetic and electric fields in plasma \cite{borghesi2002electric, borghesi2003measurement}, etc. 

A large number of studies have been performed to understand  the mechanism involved in the laser-plasma interaction-driven proton/ion acceleration. Among all possible candidates the  target normal sheath acceleration (TNSA) mechanism \cite{wilks2001energetic, snavely2000intense, mora2003plasma} has received a wider attention than other (radiation pressure-based) mechanisms.
The paramount factor has been the wide accessibility of the laser parameters required for the TNSA mechanism to operate. In this mechanism, the hot electrons generated by laser-plasma interaction at the front surface of the target escape to the rear side of the target. This electron cloud while emerging from the rear surface of the target forms a strong sheath electric field which is responsible for accelerating protons/ions to several 10s of MeV energies.


There has been a quest to enhance the laser energy absorption\cite{wilks2001energetic, snavely2000intense, mora2003plasma, goswami2021ponderomotive, ferri2019enhanced, klimo2011short, zou2019enhancement, andreev2011efficient} and generation of highly energetic, focused electrons at the rear side of the target, in order to obtain protons/ions with increasingly high energies and improved beam divergence. Among all the possible strategies to achieve this goal, structuring the target is the most economical one. The structured target results in an enhancement of proton/ion energies due to a change in interaction mechanism and an increase in the laser interaction area \cite{ferri2019enhanced, klimo2011short, zou2019enhancement, andreev2011efficient}. In the case of a normally incident laser pulse, the cut-off energy of the accelerated protons (ions) has recently been shown to strongly depend upon the shape of the structures on the front surface of the target \cite{khan2023enhanced}. Moreover, it has also been reported that in the case of flat targets, the oblique incidence of the laser pulse leads to an enhancement in the cut-off energies of the accelerated protons/ions \cite{ferri2020effects} which has been well understood on the basis of Brunel absorption mechanism \cite{brunel1987not}. Another proposed strategy to enhance the proton/ion cutoff energies is by splitting the main laser pulse into two less energetic pulses (with half the energy in each pulse), incident onto the target within a short time delay\cite{ferri2018proton, scott2012multi, markey2010spectral} or incident at the same time but at certain angles between them i.e. colliding laser pulse \cite{ferri2019enhanced}. 

\begin{figure}
	\includegraphics[height=4.5cm,width=0.7\textwidth]{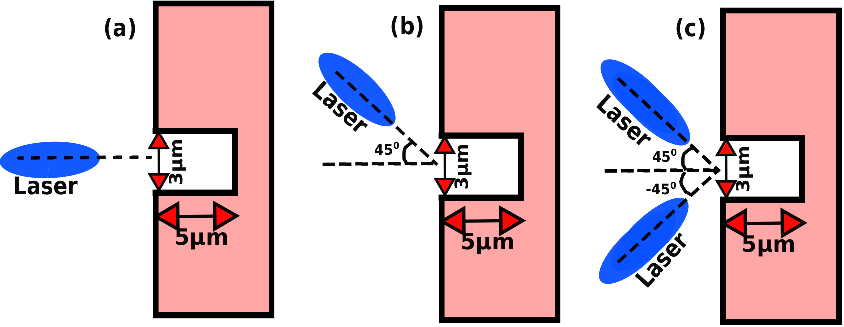}
 	\caption{\label{fig:epsart} Schematic representation of the rectangular grooved target for SNP(a), SOP(b) and TCP(c).}
  \label{fig:1}
\end{figure}

In this paper, we perform 2D PIC simulations to investigate the effect of the oblique incidence of the laser pulse on the proton/ion acceleration when a micron-sized groove is present at the front side of the target. We consider three cases, a) Single Normal Pulse (SNP): a single laser pulse incident normally at the target front surface, b) Single Oblique Pulse (SOP): a single laser pulse incident obliquely at the target front surface, and c) Two Colliding Pulse (TCP): the above single laser pulse is split into two oblique pulses of equal energy by halving the intensity and keeping all the other parameters exactly the same. These three configurations have been schematically depicted in Fig.\ref{fig:1}. 

In Sec. II, the details of the simulation setup have been provided. The 2D PIC simulation findings for the flat and grooved targets have been discussed in sections III and IV, respectively. In Sec. V, we briefly discuss the simulation results for a periodically structured (grooved) target for the completeness of our results. Finally, in Sec. VI, important outcomes of the present investigation have been summarized. 

\section{Simulation Setup}
For our simulations,  we use a two-dimensional setup of fully relativistic particle-in-cell (PIC) code, EPOCH\cite{arber2015contemporary}. The two-dimensional PIC simulations used here, in spite of over-estimating the energies of the accelerated particles \cite{sgattoni2012laser, d2013optimization,stark2017effects}, can address the salient physics of the acceleration mechanism with much limited resources as compared to full 3D simulations. The simulation domain is 90$\mu$m  along the x-axis (extends from -10$\mu$m to 80$\mu$m along the direction of laser propagation) and 189$\mu$m along the y-axis (extends between $\pm$ 94.5$\mu$m). The mesh size is 9 $\times$ 27 $\mu m^{2}$. Simple laser and open boundaries are used at the left and the right ends of the simulation box, respectively. In the transverse direction, periodic boundaries are used for fields whereas thermal boundaries are used for particles. The laser pulse considered in our simulations is p-polarised,  has a wavelength of $0.8\mu$m and intensity of $5.5\times 10^{20}$  W/cm $^{2}$. The Gaussian profile is used both in space and time, with the pulse duration(FWHM) =40$fs$ and the focal spot at the waist $ w_0 = 3\mu$m. These laser parameters are similar to the GEMINI laser at Rutherford Appleton Lab (RAL), STFC, UK\cite{scullion2017polarization}. Also, in the previous oblique incidence studies \cite{ferri2019enhanced,yang2016high}, the optimized angle of laser pulse incidence for the laser pulse was reported to lie between 30$^{\circ}$ - 45$^{\circ}$. In our present simulations, we keep the angle of incidence as +45$^{\circ}$ for a single oblique pulse case. In the case of two colliding pulses, the two laser pulses, of half the intensity each, are incident, at the target front surface, at angles $\pm $45$^{\circ}$. Moreover, to avoid confusion, the cases of single normally incident pulse (SNP), single obliquely incident pulse (SOP), and two obliquely incident pulses (TCP) are represented by solid, dashed, and dotted lines, respectively, throughout the manuscript.

A fully ionized polyethylene [(C$_2$H$_4$)$_n$] target has been used with a mass density $\rho = 0.93 g/cm^3 $ which corresponds to the number density of carbon ions, protons, and electrons as $4 \times10^{22}$ cm$^{-3}$, $8 \times10^{22}$ cm$^{-3}$, \& $3.2 \times10^{23}$ cm$^{-3}$, respectively. The number of macro particles per cell for carbon ions is chosen as 20 while 60 macro particles are used per cell each for electrons and protons. The target is localized between $\pm$94$\mu$m along the y-axis and 0 to 7$\mu$m along the x-axis. For all the simulations performed with a grooved target, a rectangular groove of optimized dimensions, groove width of 3$\mu$m, and groove depth of 5$\mu$m, as reported in \cite{khan2023enhanced} for normal incidence case, is used.

\section{Oblique incidence of a laser pulse on a flat target}
\begin{figure}
	\includegraphics[height=5cm,width=01\textwidth]{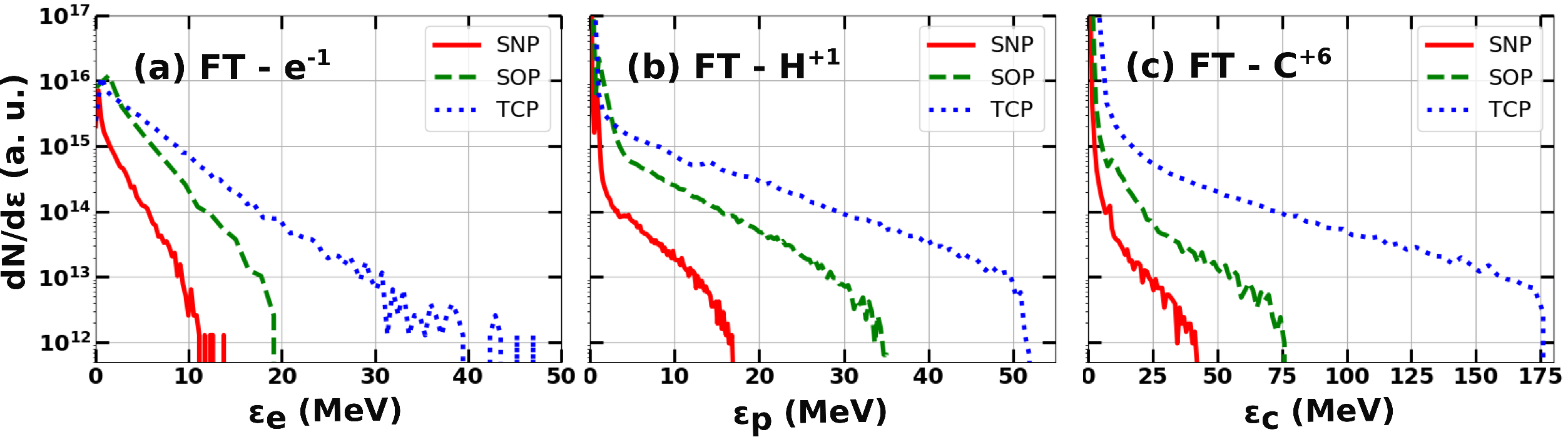}
 \caption{ The energy spectra of (a) electrons, (b) protons, and (c) Carbon ions for a flat target. The electron energy distribution is at $t=120fs$ and that for carbon ion $\&$ proton is at $t=700fs$.}
 \label{fig:2}
\end{figure}

Before discussing the results for the structured target, we first reproduce the case of a flat target irradiated by a high-intensity laser pulse. It is known that in the case of a flat target, as the angle of incidence is increased from 0 degrees (normal incidence) to 45 degrees, there is an enhancement in the laser energy absorption by electrons \cite{brunel1987not,yogo2015ion} which leads to a relatively stronger sheath electric field at the rear surface of the target. It results in an enhanced energy cutoff for protons and ions accelerated by the sheath electric field. In figure.\ref{fig:2} the simulated energy spectra of electrons, protons, and carbon ions are shown for three different laser pulse configurations, namely, a single normally incident laser pulse, a single oblique laser pulse incident at $45^\circ$ to the normal, and two oblique laser pulses (each with half the intensity) at $\pm 45^\circ$ which we term as ``two colliding pulses". The cutoff energies for electrons, protons, and carbon ions are higher in the case of oblique incidence than for the normal incidence, which is consistent with the previous numerical \cite{klimo2011short, ferri2020effects} as well as experimental \cite{ceccotti2013evidence,floquet2013micro} observations. In our simulations, we observed that there is an enhancement of proton cut-off energy roughly by a factor of two (from 17MeV to 35.2MeV) in a single oblique pulse configuration. Moreover, the cutoff energies for the three species are the highest in the case of two colliding pulses configuration in agreement with existing literature\cite{rahman2021particle, ferri2019enhanced, ferri2020effects, yao2022optimizing}.

\begin{figure}
	\includegraphics[height=5cm,width=0.8\textwidth]{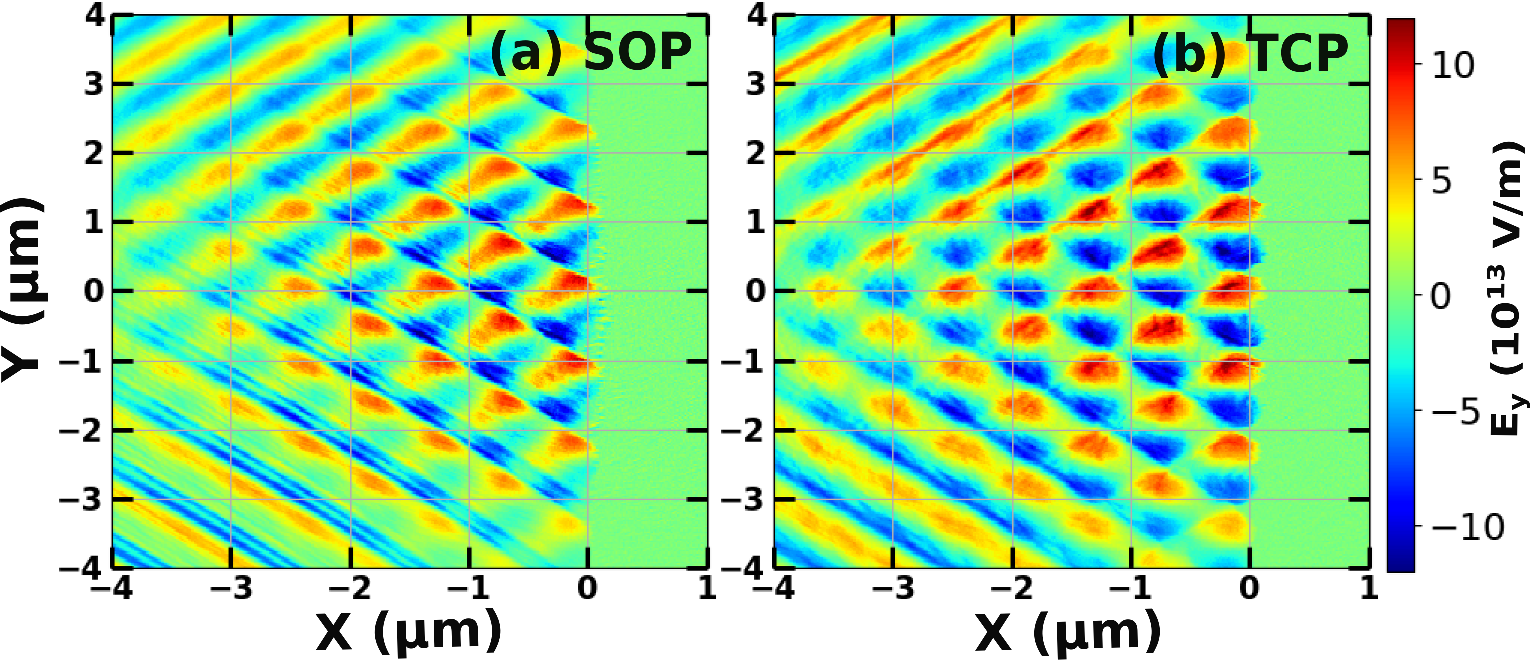}
 \caption{Interference pattern at the front side of flat tatget for SOP(left) and TCP (right)  at $t=90fs$.}
 \label{fig:3}
\end{figure}

In TCP configuration, the two pulses constructively interfere before interacting with the target \cite{rahman2021particle, ferri2019enhanced} and the peak resultant field is $\sqrt{2}$ times larger as compared to the case of a single oblique laser pulse of double the intensity \cite{ferri2019enhanced}, see figure \ref{fig:3}. 
This results in the generation of a large number of hot electrons with significantly higher temperatures. We observe in our 2D simulations that the proton cut-off energy in TCP configuration is enhanced by a factor of three compared to a single normally incident laser pulse (from 17MeV to 52.2MeV). The trend of these results is in agreement with previous studies \cite{rahman2021particle,ferri2019enhanced,yang2016high}. Moreover, the TCP configuration also maintains symmetry and its experimental implementation is easy as the pulse is not reflected back to damage the optics \cite{ferri2019enhanced, yao2022optimizing}. 

Now, a comparative study of the above three configurations (SNP, SOP $\&$ TCP) with a micron-sized grooved target is carried out to investigate if a similar trend is followed as in the case of a flat target.

\section{Oblique incidence of a laser pulse on a grooved target}

\begin{figure}
	\includegraphics[height=5cm,width=01\textwidth]{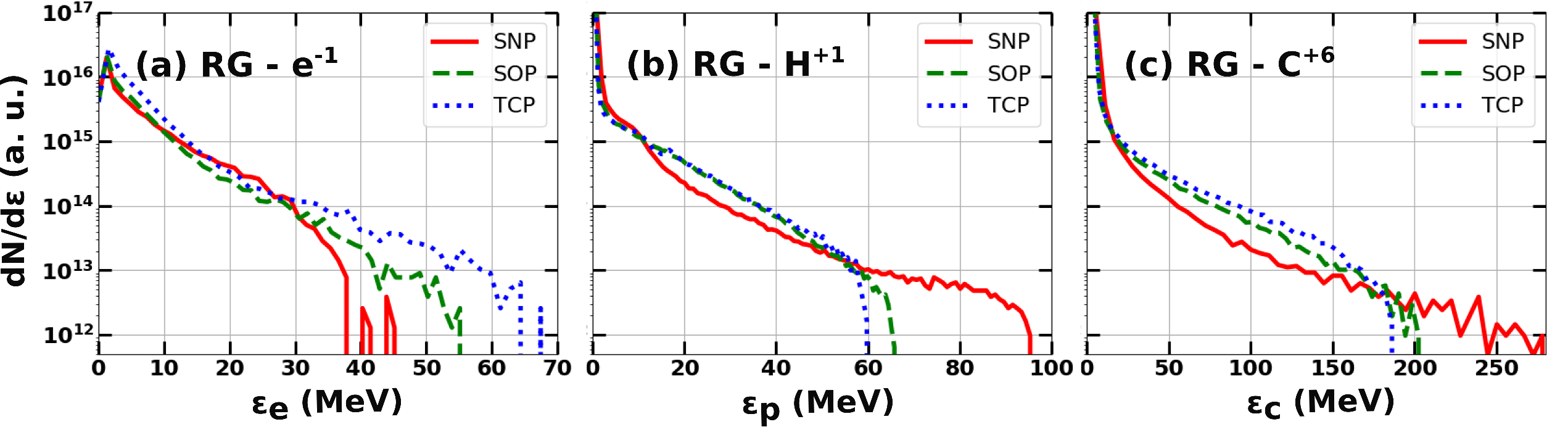}
 \caption{ The energy spectra of (a) electrons, (b) protons, and (c) Carbon ions for a target with a rectangular groove. The electron energy distribution is at $t=120fs$ and the energy distribution for carbon ions $\&$ protons are both at $t=700fs$.}
 \label{fig:4}
\end{figure}

In Fig.4 the energy spectra of electrons, protons, and carbon ions are shown for a target with a rectangular groove (RG) on its front surface, for three different laser configurations. It is worth noting that although the electron energy spectra for the RG target (Fig.\ref{fig:4}(a)) are similar to those obtained for the flat target (Fig.\ref{fig:2}(a)) the energy spectra for protons (Fig.\ref{fig:4}(b)) and carbon ions (Fig. \ref{fig:4}(c)) display entirely opposite behavior as compared to the flat target case (Fig.\ref{fig:2}(b) and Fig.\ref{fig:2}(c)). In the case of the (rectangularly) grooved target, the maximum cutoff energy happens for the SNP configuration and the minimum cutoff energy occurs for TCP configuration. 

\begin{figure}
	\includegraphics[height=4cm,width=0.8\textwidth]{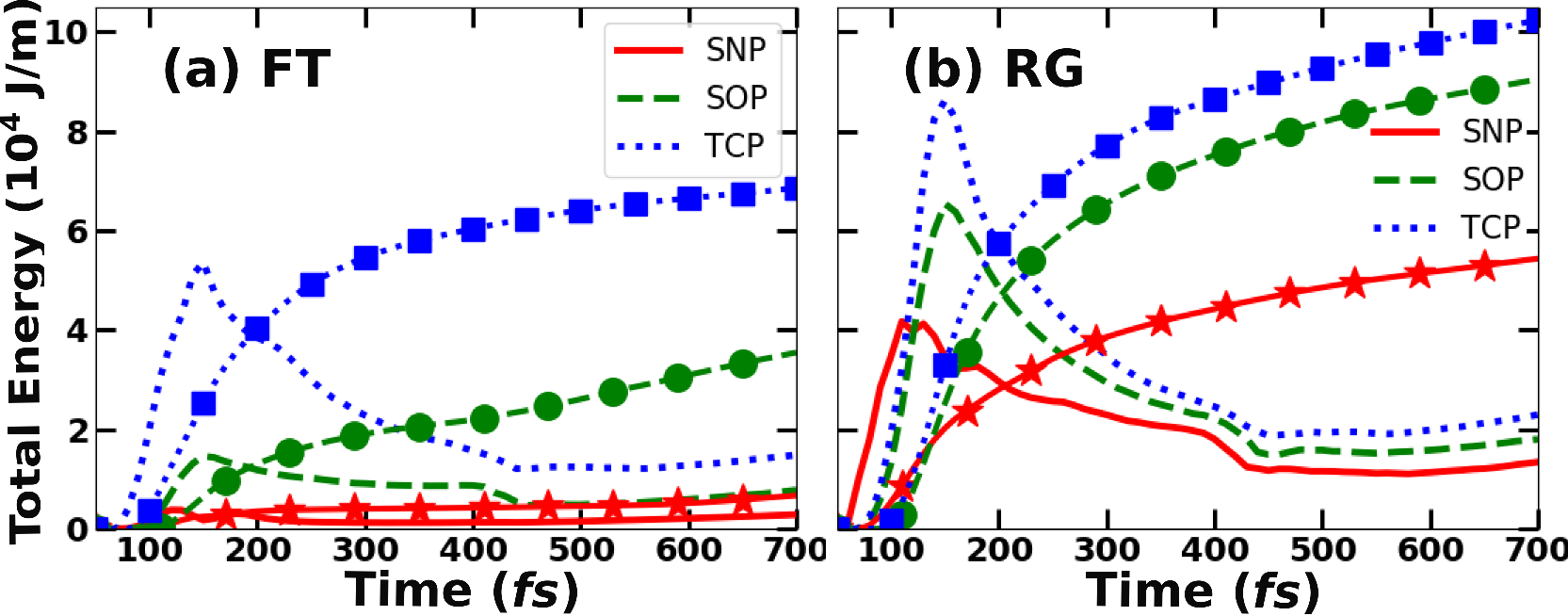}
    \caption{Time evolution of the total kinetic energy of electrons (without marker) and protons (with marker) for (a) the flat target, and (b) the grooved target.}
    \label{fig:5}
\end{figure}

\begin{figure}
	\includegraphics[height=5cm,width=0.8\textwidth]{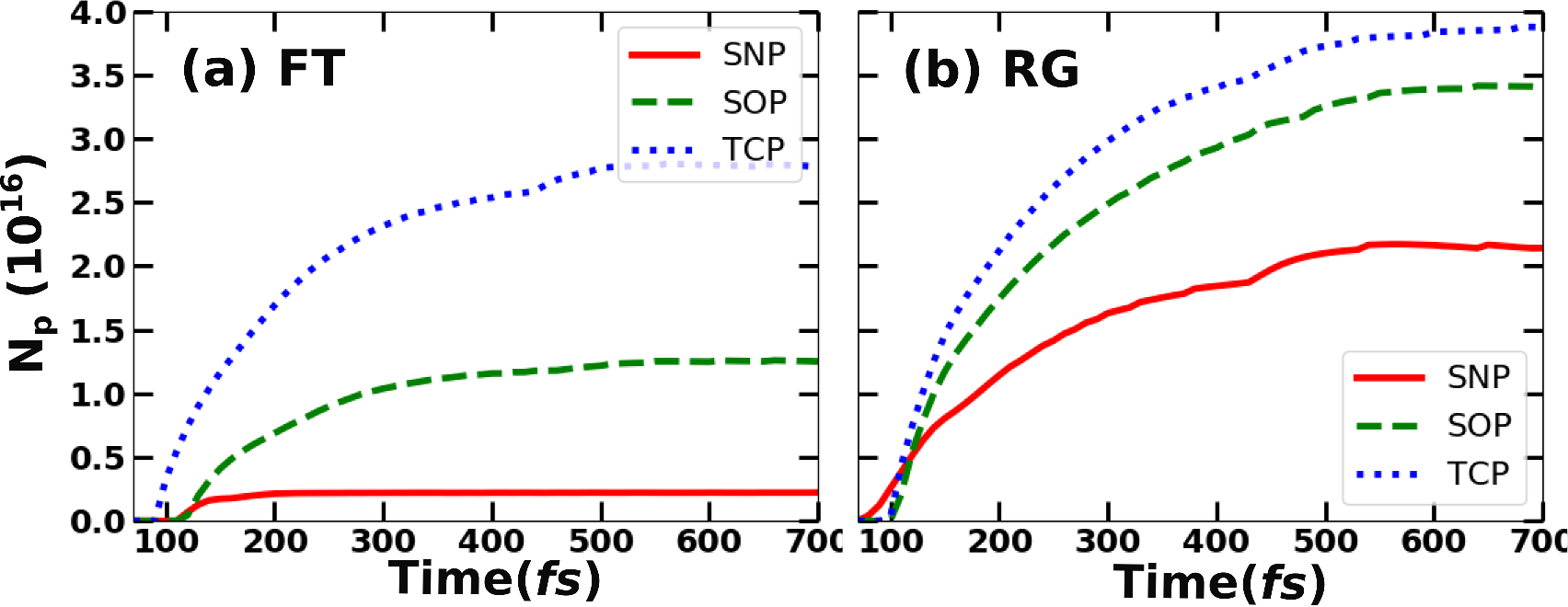}
     \caption{The Number of energetic protons moving at the rear side of the target having energy greater than 3.5 MeV for flat (left) and grooved (right) at time $t=700fs$.}
     \label{fig:6}
\end{figure}

Figure \ref{fig:5} represents the time evolution of the total kinetic energy of electrons (without marker) and protons (with marker), for the flat target (left) and for the grooved target (right). It can be seen that there is no contradiction between the results of flat and grooved target cases as far as the total energies of electrons and protons are concerned, i.e., those are highest in the TCP configuration and are lowest in the SNP configuration, for both cases. It is also observed that the number of protons at the rear side of the target (kinetic energy $\geq$ 3.5MeV) is maximum for the TCP configuration and is minimum for the SNP configuration, for both flat and grooved targets, as shown in figure \ref{fig:6}a $\&$ \ref{fig:6}b, respectively. 

So the number of energetic protons as well as their total kinetic energy is maximum in the TCP case, for both the flat and the grooved targets, but the highest proton (and ion) cut-off energy for the grooved target case is obtained in the SNP configuration unlike for the flat target case.

To understand the reason behind this change in the grooved target case, we estimate the number of protons with energy lying in the ranges 0 - 50 MeV and 50 - 100 MeV, for the different configurations, SNP, SOP, and TCP. The numbers of the protons that appear on the rear side of the target are given in Table 1. Note that the overall number of protons in the TCP configuration is maximum for both flat as well as grooved targets. Also, the number of protons having energy in the range of 0-50 MeV is maximum for the TCP configuration and minimum for the SNP configuration, for both types of targets. However, if we look at the number of protons in the energy range 50 - 100 MeV, interestingly, for the grooved target, the SNP configuration has the maximum number of protons (followed by the SOP and TCP cases), unlike in the case of the flat target. Clearly, a larger number of highly energetic protons is generated in the normal incidence case when the target has a micron-sized groove on its front surface.

\begin{table}[ht]
\centering
\caption{Number of protons at the rear side of the target with a rectangular groove, at time $t=700fs$.}

\begin{tabular}{|c| c| c| c|} 
 \hline
$\varepsilon_p$ (MeV) & \multicolumn{3}{c|}{$N_p$ (1e14)} \\ 
\cline{2-4}  & SNP & SOP & TCP \\ 
 \hline
 0-50 & 8505 & 10762 & 12003\\
 \hline
 50-100 & 3.70 & 2.60 & 2.64 \\ 
 \hline
\end{tabular}
\label{tab:caption}
\end{table}%

The underlying physics can be understood by looking at the laser pulse intensity distribution inside the groove structure, shown in Fig.7. Apparently, there is an enhancement of laser pulse intensity by redistribution, as it passes through the aperture\cite{ji2016towards}. The level of intensification is given by $\eta=I/ I_0$ where $I_0 \: \& \: I$ are the intensities in the vacuum and in the presence of the target. The magnitude of the intensification is $\eta = 7.97,\: 5.26,\:  \& \: 5.12$ for SNP, SOP $\&$ TCP respectively in RG target at time $t=90fs$. The proton cutoff energy is in accordance with the laser pulse intensification. \\

\begin{figure}
	\includegraphics[height=5cm,width=01\textwidth]{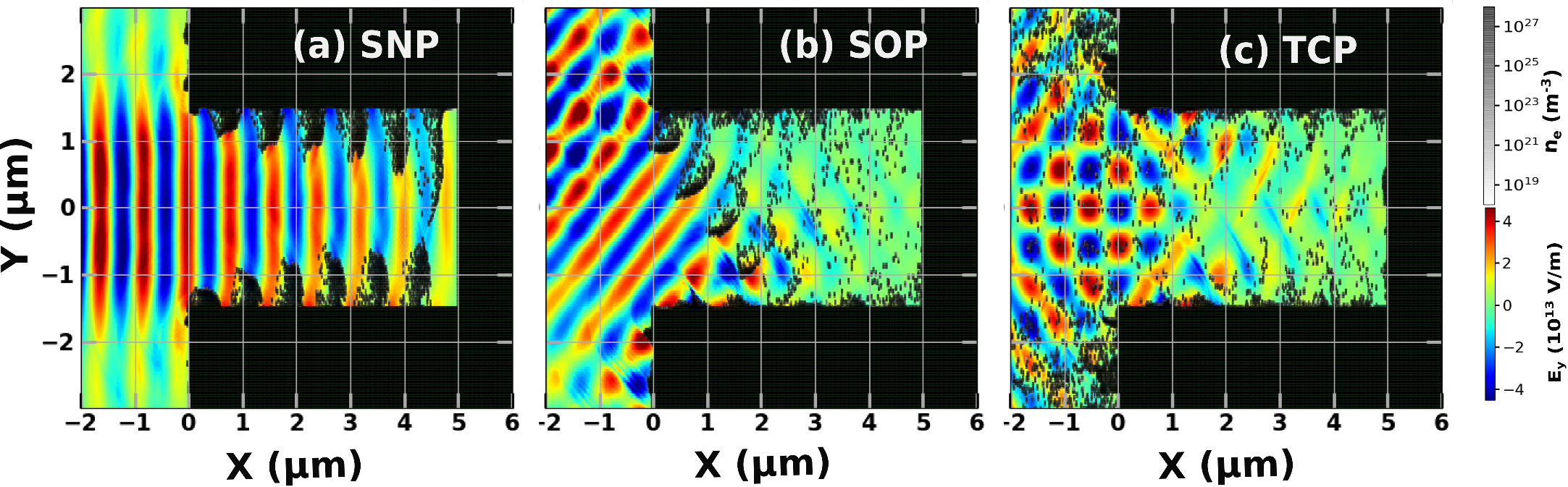}
        \caption{Spatial distribution of electron number density and laser electric field($E_y$) for (a) SNP at $t=50fs$, (b) SOP at $t=60fs$ and TCP at $t=60fs$. }
        \label{fig:7}
\end{figure}

\begin{figure}
	\includegraphics[height=9cm,width=01\textwidth]{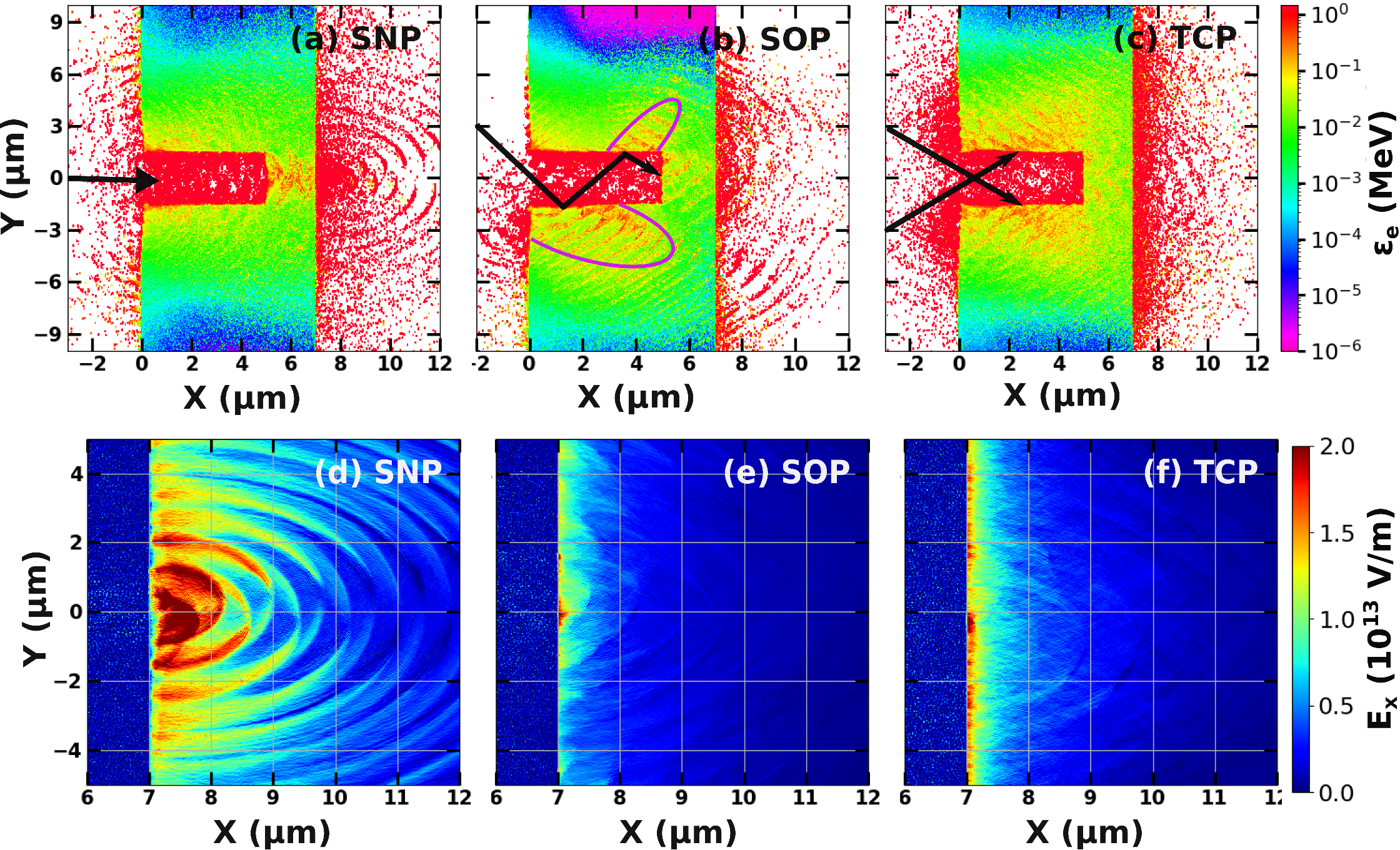}
        \caption{Electron energy distribution(first row) in xy-plane for the rectangular grooved target SNP (a), SOP (b), and TCP (c) at time $t=90~fs$. The second row shows the corresponding rear side sheath field at the same time.}
        \label{fig:8}
\end{figure}

For SNP, the electromagnetic field intensity is enhanced by more than six times inside the groove (Fig.\ref{fig:7}(a)). It is found that as the pulse is perfectly aligned along the central axis of the rectangular groove, during every half period of the laser the transverse component of the laser electric field ($E_{y}$) pulls out a large number of electron bunches from the top and bottom side walls of the groove \cite{khan2023enhanced}. A quite similar type of electron bunching and acceleration has also been reported by Snyder et al.\cite{snyder2019relativistic} and Zhu et al.\cite{zhu2022bunched} in which they use a circular micro-channel plate and a metal cone with a thin foil at its apex. These hot electrons are trapped and phase-locked with the laser pulse and are accelerated in the forward direction as the trains of high-density bunches of electrons. As the target is over-dense, the laser pulse is reflected back and the hot electron bunches decouple from the laser and come out at the rear side of the target as can be seen in Fig.\ref{fig:8}(a). These high-density hot electrons are highly focused and form a strong electrostatic sheath field at the rear side of the target (Fig.\ref{fig:8}(d)) resulting in a relatively higher cut-off energy of protons.\\

 For SOP (Fig.\ref{fig:7}(b)), when the laser pulse is incident obliquely from the upper half of the incidence plane, it first interacts with the lower side wall of the groove. The transverse component of the electromagnetic field ($E_y$) partially interacts with the front edge of the upper side wall and extracts electron bunches. These hot electron bunches, along with the laser pulse interact with the lower side wall of the groove. The laser pulse interacting with the lower side wall of the groove transfers a significant portion of its energy there and gets reflected towards the upper side wall with reduced energy. The pulse energy transferred to the upper side wall is therefore much less. A large number of energetic electrons are generated at the lower wall whereas a relatively lower number of energetic electrons are generated at the upper wall of the groove are clearly visible in the spatial electron energy distribution in Fig.\ref{fig:8}(b). These hot electrons generated at the lower and upper sidewalls travel within the target toward the rear side and emerge from the rear surface in two different directions. Thus they result in a diverging (asymmetric) sheath field (Fig. \ref{fig:8}(e)) leading to a decrease in the cut-off energy of the protons as compared to the SNP configuration. Also the incident and reflected part of the laser pulse form an interference pattern near the lower and upper side walls of the groove in a way that the intensification value ($\eta =5.26$) is less than that observed in the SNP configuration ($\eta =7.97$) and is localized only at certain points thus explaining the reduced acceleration of protons in this case. \\

\begin{figure}
	\includegraphics[height=4.5cm,width=0.6\textwidth]{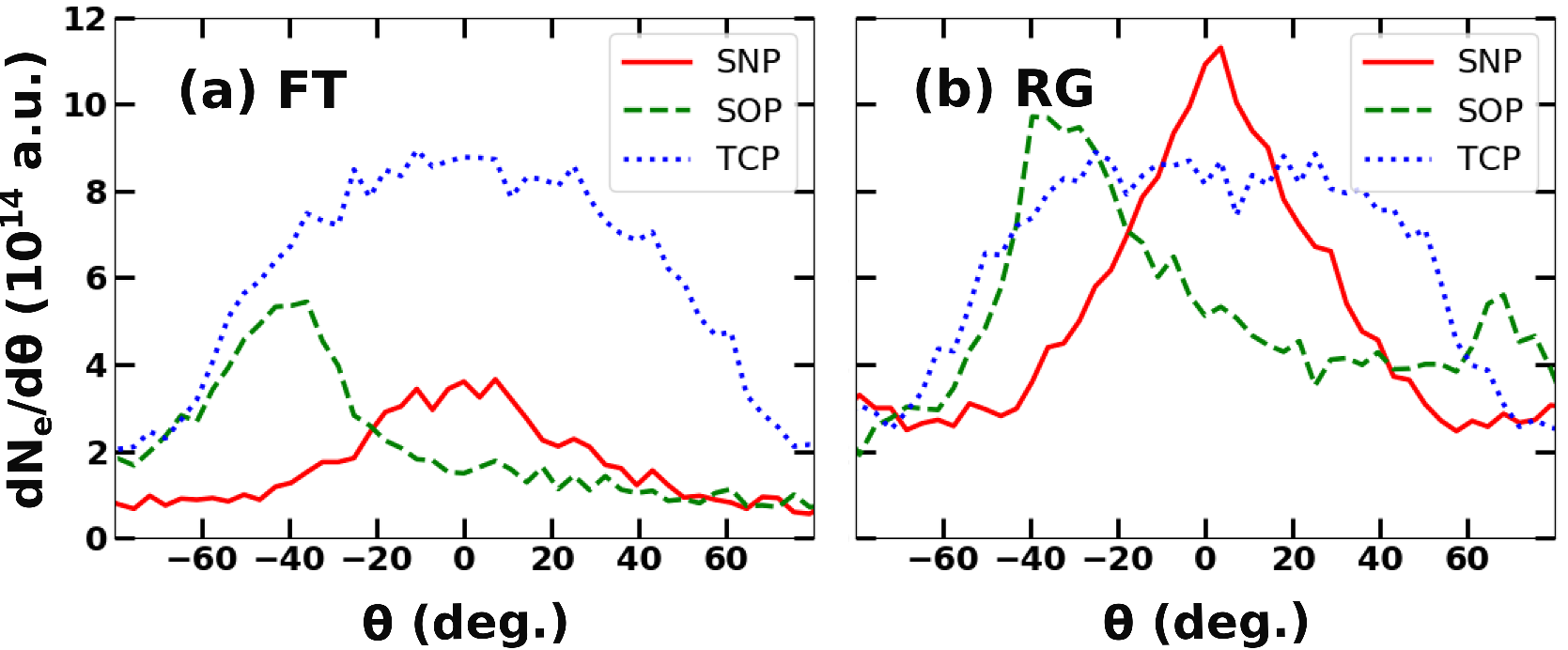}
        \caption{Electron energy angular distribution for flat (left) and rectangular grooved target at $t=11 0fs$.}
        \label{fig:9}
\end{figure}

\begin{figure}
	\includegraphics[height=4.5cm,width=0.4\textwidth]{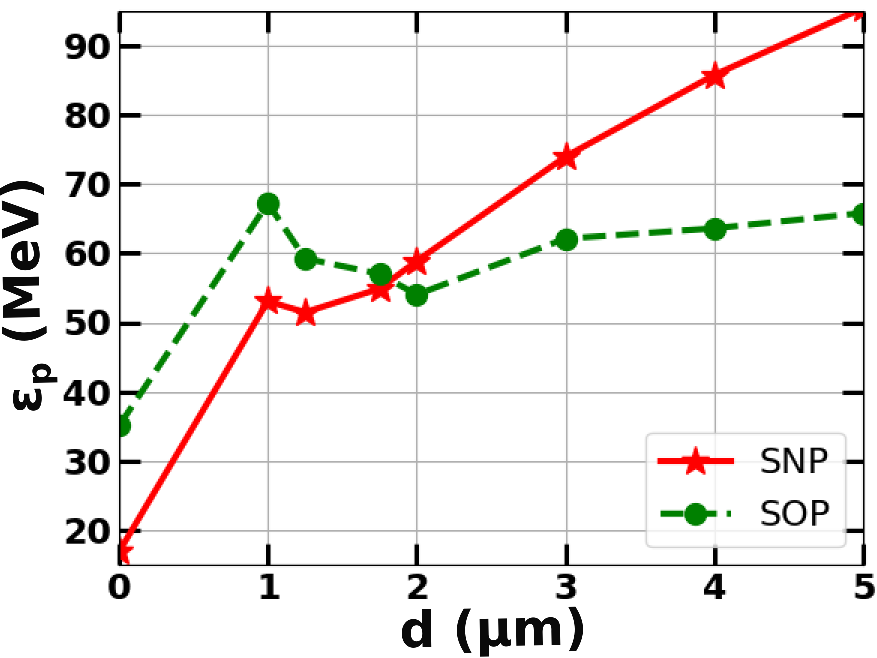}
        \caption{Proton energy cutoff energy for different groove depth keeping the width constant (3$\mu m$), at $t=700fs$. The solid(red) line is for SNP and the dashed (green) line is for SOP. }
        \label{fig:10}
\end{figure}

\begin{figure}
	\includegraphics[height=4.5cm,width=0.8\textwidth]{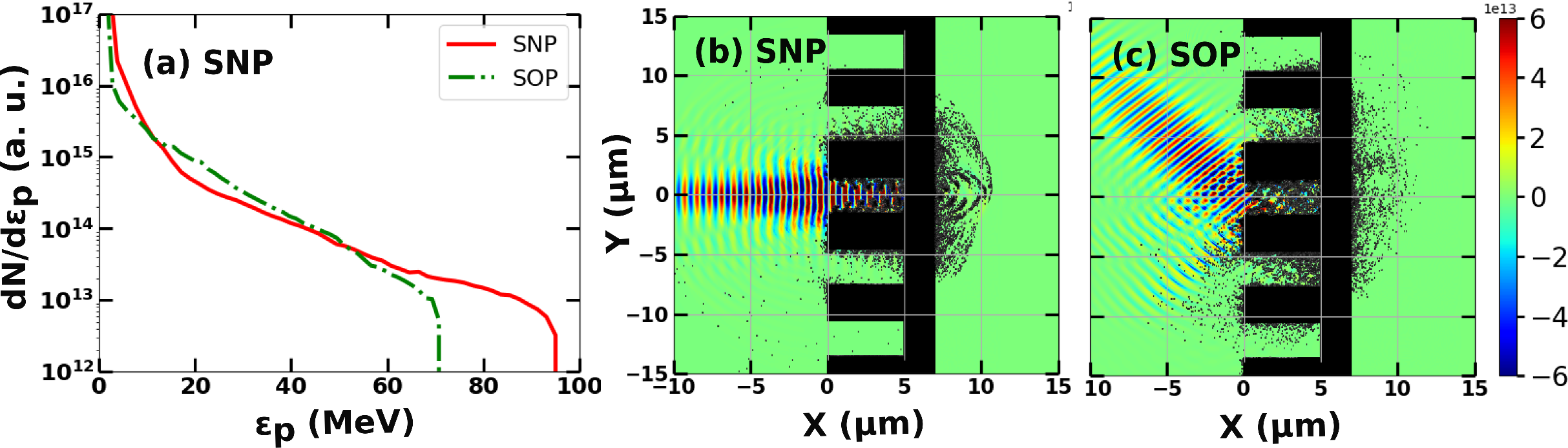}
        \caption{(a) Proton energy distribution for the periodic grooves for SNP (solid) and SOP (dashed). Spatial distribution of electron number density and laser electric field($E_y$) for (b) SNP at $t=70fs$, (c) SOP at $t=80fs$. The groove width, depth, and period are $3\mu$m and $5\mu$m, and $6\mu$m respectively. }
        \label{fig:11}
\end{figure}

\begin{figure}
	\includegraphics[height=4.5cm,width=0.4\textwidth]{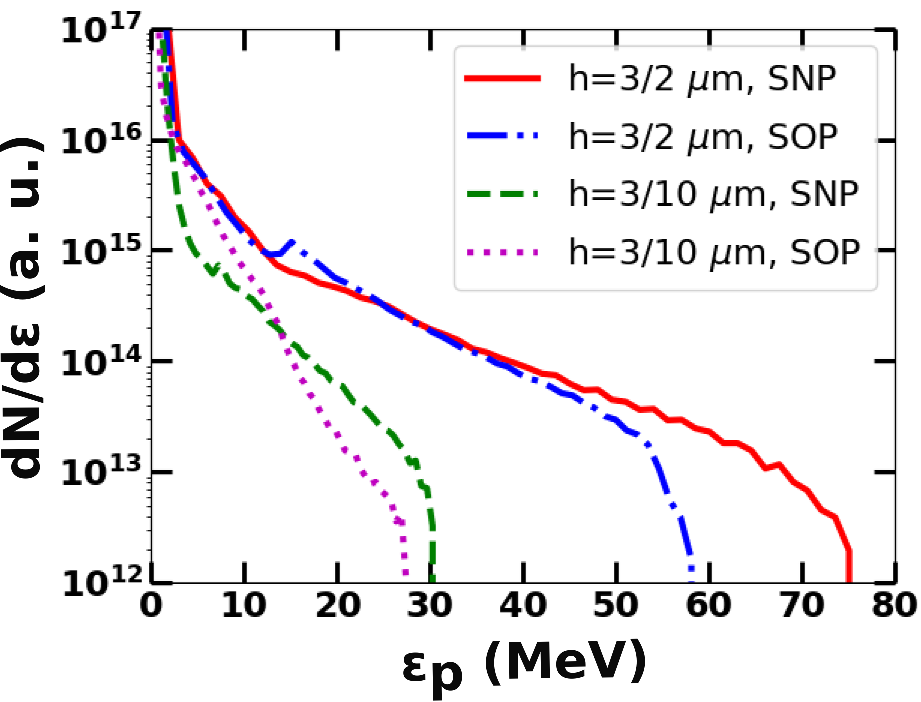}
        \caption{Effect of the width (and separation) of the grooves on the proton energy distribution for the SNP and the SOP cases at time $t=550fs$. The separation between two consecutive grooves is same as the groove width.}
        \label{fig:12}
\end{figure}
 In TCP configuration (Fig.\ref{fig:7}(c)), when two pulses collide, they form an interference pattern before entering the groove. The modified intensity is distributed such that the intensity maxima at the longitudinal axis of the groove are much higher than in the proximity of the side walls. Hence the energy transfer to the side wall electrons is much less than in the SNP configuration. The low energy electrons generated at both the side walls of the groove move inside the target toward the rear side with an angle between them like in the SOP configuration. However, in this case, the electron density distribution is symmetric in nature as the energy transfer by the laser pulse to the upper and lower side wall electrons is equal as can be seen in Fig.\ref{fig:8}(c). Due to the interference pattern formed by the colliding lasers, there are no high-energy electrons generated from the front edges of the groove as in the SOP configuration. Hence, the electrons are comparatively less energetic and have the highest divergence (Fig.\ref{fig:8}(f)). These diverging electrons form a very weak accelerating sheath field leading to reduced cut-off energy of protons as well as of ions.

This is in contrast with the flat target case (Fig.\ref{fig:9}(a)), for which the patterns of the angular distribution for the SNP and TCP configurations are similar but the number of electrons is much higher in the TCP configuration. This leads to the formation of a stronger sheath field at the rear side in the TCP case. For the rectangular groove target, the electrons are widely distributed in TCP and highly focused in SNP (Fig.\ref{fig:9}(b)) which results in a relatively stronger sheath formation at the rear side of the target in the SNP configuration. The SOP results in the number of protons in between that of SNP $\&$ TCP but the distribution is asymmetric in nature. 

On the basis of our simulation results it can be stated that for the fixed width and depth of the rectangular groove, the SNP configuration is more effective than SOP and TCP configurations. We further studied the effect of groove depth on the protons' cutoff energy. It can be seen from Fig.\ref{fig:10}, that there is a continuous increase in proton cut of energy with an increase in the groove depth while keeping the groove width fixed (3$\mu m$), in the SNP configuration. This has also been reported in our earlier work \cite{khan2023enhanced}. Now, for the SOP configuration,  the proton cutoff energy shows a drastic change in its dependence on the groove depth. For the small groove depth (d=1$\mu$m), the SOP has greater cutoff energy than SNP, and beyond d=2$\mu$m, the SOP configuration results in almost constant cutoff energy which remains less than the cutoff energy obtained in the SNP configuration.  

There is a small dip in the cutoff energy around d=1.5$\mu$m in the SOP configuration possibly due to the following reasons. Since the laser pulse is incident at a 45-degree angle, the centre of its front interacts with the lower wall of the groove at x = 1.5$\mu$m. When the groove's depth is smaller than 1.5$\mu$m, more than half of the laser pulse's cross-section interacts with the groove's front wall, resulting in more effective generation of hot electrons (most probably due to Brunel's mechanism) which then move towards the rear side in the longitudinal direction exiting the rear surface of the target along its normal. This results in a stronger sheath field, and hence a higher proton energy cutoff, than in the SNP configuration. For the groove depth of more than 1.5$\mu$m, more than half of the laser pulse's cross-section interacts with the groove's lower side wall, resulting in electron motion primarily in the oblique direction, leading to a relatively weaker sheath field.  

\section{Oblique incidence of a laser pulse on a periodically structured target}
Our investigations, discussed in Sec III and Sec IV, indicate that while for a flat TNSA target, an obliquely incident laser pulse is more effective for generating high-energy protons as compared to a normally incident laser pulse, in the case of a target with a single rectangular groove at its front surface, the normally incident laser pulse results in higher cutoff energy of protons as compared to the oblique incidence case. Moreover, the normally incident single laser pulse is also more effective than two obliquely ($\pm 45^\circ$) incident laser pulses of half the energy (TCP configuration) which was found to be the most advantageous in the case of a flat target \cite{ferri2019enhanced, rahman2021particle}. We have also verified this behaviour for other geometries of the groove. Those results are not included here for the sake of brevity.

To examine the wider applicability of our results we have also investigated the case of a periodically structured target. The target has multiple grooves at its front surface, each of width = 3$\mu$m, and depth = 5$\mu$m. Two consecutive grooves are separated by 3 $\mu$m. The proton energy spectra for such a structured target are shown, in Fig.\ref{fig:11}(a). When compared to the SNP configuration (Fig.\ref{fig:11}(b)), in the SOP configuration (Fig.\ref{fig:11}(c)), the target's rear electron distribution is more diverged, leading to the production of a weak sheath field. As a result, the SOP configuration results in a decrease in the proton cutoff energy in comparison to the SNP configuration. 

One may intrigue how the dimensions of the periodically separated grooves on the front surface of the target affect the SNP and SOP scenarios, especially when the groove width is much narrower than the laser beam diameter thereby not allowing the laser pulse to completely get into the groove in the normal incidence case while mostly getting reflected for the oblique incidence. 

It is observed that the proton cut off energy decreases as the width and the separation of the grooves are simultaneously reduced while maintaining their ratio constant (chosen 1 here). This has been shown in figure \ref{fig:12} for two groove widths, h=3/2 $\mu$m, and h=3/10 $\mu$m. For these groove widths  (as well as many other groove widths the simulation results for which are not included here) the proton cutoff energy for the SOP case is significantly lower than the SNP case.
These results further support our conclusion that the normally incident laser pulse is more effective for the generation of energetic protons in the case of a structured target. This is an important result given the completely opposite knowledge in the case of flat targets.

\section{Conclusions}
We have investigated, using two-dimensional PIC simulations, the effect of different configurations, namely, a single laser pulse at normal incidence (SNP), a single laser pulse at oblique ($45^\circ$) incidence (SOP), and two laser pulses (of half the intensity) at oblique incidence ($\pm 45\circ$) (TCP), on the micron-sized groove at the front surface of a hydrocarbon target, in the TNSA regime. There is an enhancement of the cut-off energy for the grooved target as compared to the flat target in all three cases. For the flat target SNP configuration results in a minimum cutoff energy of protons as well as Carbon ions while the TCP configuration results in a maximum cut-off energy of both species. This trend gets reversed for the (rectangularly) grooved target and the SNP configuration is found to be the most effective one for proton as well as ion acceleration. This holds equally good for other geometries of the groove. To further strengthen our claim, we have also verified this change of trend in the cutoff energies ($E_{SNP} > E_{SOP}$) for a periodically micro-structured target. From our results, we can conclude that although the oblique incidence is more effective for proton acceleration from laser-irradiated flat targets, in the case of targets with structured (with a single groove or multiple periodically spaced grooves) front surface, a normal incidence of the laser pulse results in a higher cutoff energy of protons (as well as of ions) as compared to the oblique incidence.

\begin{acknowledgments}
The authors would like to acknowledge the EPOCH consortium, for providing access to the EPOCH-4.9.0 framework \cite{arber2015contemporary}, and high-performance computing (HPC) facility at the Indian Institute of Technology Delhi for computational resources. IK also acknowledges the University Grants Commission (UGC), govt. of India, for his senior research fellowship (Grant no. 1306/(CSIR-UGC NET DEC. 2018)). Moreover, the constructive feedback on the manuscript from Andrea Macchi of CNR, University of Pisa, Italy, is gratefully acknowledged.   
\end{acknowledgments}
\section*{Data Availability}
The data that support the findings of this study are available from the authors upon reasonable request. 

\begin{thebibliography}{36}%
\makeatletter
\providecommand \@ifxundefined [1]{%
 \@ifx{#1\undefined}
}%
\providecommand \@ifnum [1]{%
 \ifnum #1\expandafter \@firstoftwo
 \else \expandafter \@secondoftwo
 \fi
}%
\providecommand \@ifx [1]{%
 \ifx #1\expandafter \@firstoftwo
 \else \expandafter \@secondoftwo
 \fi
}%
\providecommand \natexlab [1]{#1}%
\providecommand \enquote  [1]{``#1''}%
\providecommand \bibnamefont  [1]{#1}%
\providecommand \bibfnamefont [1]{#1}%
\providecommand \citenamefont [1]{#1}%
\providecommand \href@noop [0]{\@secondoftwo}%
\providecommand \href [0]{\begingroup \@sanitize@url \@href}%
\providecommand \@href[1]{\@@startlink{#1}\@@href}%
\providecommand \@@href[1]{\endgroup#1\@@endlink}%
\providecommand \@sanitize@url [0]{\catcode `\\12\catcode `\$12\catcode
  `\&12\catcode `\#12\catcode `\^12\catcode `\_12\catcode `\%12\relax}%
\providecommand \@@startlink[1]{}%
\providecommand \@@endlink[0]{}%
\providecommand \url  [0]{\begingroup\@sanitize@url \@url }%
\providecommand \@url [1]{\endgroup\@href {#1}{\urlprefix }}%
\providecommand \urlprefix  [0]{URL }%
\providecommand \Eprint [0]{\href }%
\providecommand \doibase [0]{http://dx.doi.org/}%
\providecommand \selectlanguage [0]{\@gobble}%
\providecommand \bibinfo  [0]{\@secondoftwo}%
\providecommand \bibfield  [0]{\@secondoftwo}%
\providecommand \translation [1]{[#1]}%
\providecommand \BibitemOpen [0]{}%
\providecommand \bibitemStop [0]{}%
\providecommand \bibitemNoStop [0]{.\EOS\space}%
\providecommand \EOS [0]{\spacefactor3000\relax}%
\providecommand \BibitemShut  [1]{\csname bibitem#1\endcsname}%
\let\auto@bib@innerbib\@empty
\bibitem [{\citenamefont {Patel}\ \emph {et~al.}(2003)\citenamefont {Patel},
  \citenamefont {Mackinnon}, \citenamefont {Key}, \citenamefont {Cowan},
  \citenamefont {Foord}, \citenamefont {Allen}, \citenamefont {Price},
  \citenamefont {Ruhl}, \citenamefont {Springer},\ and\ \citenamefont
  {Stephens}}]{patel2003isochoric}%
  \BibitemOpen
  \bibfield  {author} {\bibinfo {author} {\bibfnamefont {P.}~\bibnamefont
  {Patel}}, \bibinfo {author} {\bibfnamefont {A.}~\bibnamefont {Mackinnon}},
  \bibinfo {author} {\bibfnamefont {M.}~\bibnamefont {Key}}, \bibinfo {author}
  {\bibfnamefont {T.}~\bibnamefont {Cowan}}, \bibinfo {author} {\bibfnamefont
  {M.}~\bibnamefont {Foord}}, \bibinfo {author} {\bibfnamefont
  {M.}~\bibnamefont {Allen}}, \bibinfo {author} {\bibfnamefont
  {D.}~\bibnamefont {Price}}, \bibinfo {author} {\bibfnamefont
  {H.}~\bibnamefont {Ruhl}}, \bibinfo {author} {\bibfnamefont {P.}~\bibnamefont
  {Springer}}, \ and\ \bibinfo {author} {\bibfnamefont {R.}~\bibnamefont
  {Stephens}},\ }\bibfield  {title} {\enquote {\bibinfo {title} {Isochoric
  heating of solid-density matter with an ultrafast proton beam},}\ }\href@noop
  {} {\bibfield  {journal} {\bibinfo  {journal} {Physical review letters}\
  }\textbf {\bibinfo {volume} {91}},\ \bibinfo {pages} {125004} (\bibinfo
  {year} {2003})}\BibitemShut {NoStop}%
\bibitem [{\citenamefont {Mackinnon}\ \emph {et~al.}(2006)\citenamefont
  {Mackinnon}, \citenamefont {Patel}, \citenamefont {Borghesi}, \citenamefont
  {Clarke}, \citenamefont {Freeman}, \citenamefont {Habara}, \citenamefont
  {Hatchett}, \citenamefont {Hey}, \citenamefont {Hicks}, \citenamefont {Kar}
  \emph {et~al.}}]{mackinnon2006proton}%
  \BibitemOpen
  \bibfield  {author} {\bibinfo {author} {\bibfnamefont {A.}~\bibnamefont
  {Mackinnon}}, \bibinfo {author} {\bibfnamefont {P.}~\bibnamefont {Patel}},
  \bibinfo {author} {\bibfnamefont {M.}~\bibnamefont {Borghesi}}, \bibinfo
  {author} {\bibfnamefont {R.}~\bibnamefont {Clarke}}, \bibinfo {author}
  {\bibfnamefont {R.}~\bibnamefont {Freeman}}, \bibinfo {author} {\bibfnamefont
  {H.}~\bibnamefont {Habara}}, \bibinfo {author} {\bibfnamefont
  {S.}~\bibnamefont {Hatchett}}, \bibinfo {author} {\bibfnamefont
  {D.}~\bibnamefont {Hey}}, \bibinfo {author} {\bibfnamefont {D.}~\bibnamefont
  {Hicks}}, \bibinfo {author} {\bibfnamefont {S.}~\bibnamefont {Kar}},  \emph
  {et~al.},\ }\bibfield  {title} {\enquote {\bibinfo {title} {Proton
  radiography of a laser-driven implosion},}\ }\href@noop {} {\bibfield
  {journal} {\bibinfo  {journal} {Physical review letters}\ }\textbf {\bibinfo
  {volume} {97}},\ \bibinfo {pages} {045001} (\bibinfo {year}
  {2006})}\BibitemShut {NoStop}%
\bibitem [{\citenamefont {Bulanov}\ and\ \citenamefont
  {Khoroshkov}(2002)}]{bulanov2002feasibility}%
  \BibitemOpen
  \bibfield  {author} {\bibinfo {author} {\bibfnamefont {S.}~\bibnamefont
  {Bulanov}}\ and\ \bibinfo {author} {\bibfnamefont {V.}~\bibnamefont
  {Khoroshkov}},\ }\bibfield  {title} {\enquote {\bibinfo {title} {Feasibility
  of using laser ion accelerators in proton therapy},}\ }\href@noop {}
  {\bibfield  {journal} {\bibinfo  {journal} {Plasma Physics Reports}\ }\textbf
  {\bibinfo {volume} {28}},\ \bibinfo {pages} {453--456} (\bibinfo {year}
  {2002})}\BibitemShut {NoStop}%
\bibitem [{\citenamefont {Ledingham}\ \emph {et~al.}(2014)\citenamefont
  {Ledingham}, \citenamefont {Bolton}, \citenamefont {Shikazono},\ and\
  \citenamefont {Ma}}]{ledingham2014towards}%
  \BibitemOpen
  \bibfield  {author} {\bibinfo {author} {\bibfnamefont {K.~W.}\ \bibnamefont
  {Ledingham}}, \bibinfo {author} {\bibfnamefont {P.~R.}\ \bibnamefont
  {Bolton}}, \bibinfo {author} {\bibfnamefont {N.}~\bibnamefont {Shikazono}}, \
  and\ \bibinfo {author} {\bibfnamefont {C.-M.~C.}\ \bibnamefont {Ma}},\
  }\bibfield  {title} {\enquote {\bibinfo {title} {Towards laser driven hadron
  cancer radiotherapy: A review of progress},}\ }\href@noop {} {\bibfield
  {journal} {\bibinfo  {journal} {Applied Sciences}\ }\textbf {\bibinfo
  {volume} {4}},\ \bibinfo {pages} {402--443} (\bibinfo {year}
  {2014})}\BibitemShut {NoStop}%
\bibitem [{\citenamefont {Roth}\ \emph {et~al.}(2001)\citenamefont {Roth},
  \citenamefont {Cowan}, \citenamefont {Key}, \citenamefont {Hatchett},
  \citenamefont {Brown}, \citenamefont {Fountain}, \citenamefont {Johnson},
  \citenamefont {Pennington}, \citenamefont {Snavely}, \citenamefont {Wilks}
  \emph {et~al.}}]{roth2001fast}%
  \BibitemOpen
  \bibfield  {author} {\bibinfo {author} {\bibfnamefont {M.}~\bibnamefont
  {Roth}}, \bibinfo {author} {\bibfnamefont {T.}~\bibnamefont {Cowan}},
  \bibinfo {author} {\bibfnamefont {M.}~\bibnamefont {Key}}, \bibinfo {author}
  {\bibfnamefont {S.}~\bibnamefont {Hatchett}}, \bibinfo {author}
  {\bibfnamefont {C.}~\bibnamefont {Brown}}, \bibinfo {author} {\bibfnamefont
  {W.}~\bibnamefont {Fountain}}, \bibinfo {author} {\bibfnamefont
  {J.}~\bibnamefont {Johnson}}, \bibinfo {author} {\bibfnamefont
  {D.}~\bibnamefont {Pennington}}, \bibinfo {author} {\bibfnamefont
  {R.}~\bibnamefont {Snavely}}, \bibinfo {author} {\bibfnamefont
  {S.}~\bibnamefont {Wilks}},  \emph {et~al.},\ }\bibfield  {title} {\enquote
  {\bibinfo {title} {Fast ignition by intense laser-accelerated proton
  beams},}\ }\href@noop {} {\bibfield  {journal} {\bibinfo  {journal} {Physical
  review letters}\ }\textbf {\bibinfo {volume} {86}},\ \bibinfo {pages} {436}
  (\bibinfo {year} {2001})}\BibitemShut {NoStop}%
\bibitem [{\citenamefont {Atzeni}, \citenamefont {Temporal},\ and\
  \citenamefont {Honrubia}(2002)}]{atzeni2002first}%
  \BibitemOpen
  \bibfield  {author} {\bibinfo {author} {\bibfnamefont {S.}~\bibnamefont
  {Atzeni}}, \bibinfo {author} {\bibfnamefont {M.}~\bibnamefont {Temporal}}, \
  and\ \bibinfo {author} {\bibfnamefont {J.}~\bibnamefont {Honrubia}},\
  }\bibfield  {title} {\enquote {\bibinfo {title} {A first analysis of fast
  ignition of precompressed icf fuel by laser-accelerated protons},}\
  }\href@noop {} {\bibfield  {journal} {\bibinfo  {journal} {Nuclear fusion}\
  }\textbf {\bibinfo {volume} {42}},\ \bibinfo {pages} {L1} (\bibinfo {year}
  {2002})}\BibitemShut {NoStop}%
\bibitem [{\citenamefont {Borghesi}\ \emph {et~al.}(2002)\citenamefont
  {Borghesi}, \citenamefont {Campbell}, \citenamefont {Schiavi}, \citenamefont
  {Haines}, \citenamefont {Willi}, \citenamefont {MacKinnon}, \citenamefont
  {Patel}, \citenamefont {Gizzi}, \citenamefont {Galimberti}, \citenamefont
  {Clarke} \emph {et~al.}}]{borghesi2002electric}%
  \BibitemOpen
  \bibfield  {author} {\bibinfo {author} {\bibfnamefont {M.}~\bibnamefont
  {Borghesi}}, \bibinfo {author} {\bibfnamefont {D.}~\bibnamefont {Campbell}},
  \bibinfo {author} {\bibfnamefont {A.}~\bibnamefont {Schiavi}}, \bibinfo
  {author} {\bibfnamefont {M.}~\bibnamefont {Haines}}, \bibinfo {author}
  {\bibfnamefont {O.}~\bibnamefont {Willi}}, \bibinfo {author} {\bibfnamefont
  {A.}~\bibnamefont {MacKinnon}}, \bibinfo {author} {\bibfnamefont
  {P.}~\bibnamefont {Patel}}, \bibinfo {author} {\bibfnamefont
  {L.}~\bibnamefont {Gizzi}}, \bibinfo {author} {\bibfnamefont
  {M.}~\bibnamefont {Galimberti}}, \bibinfo {author} {\bibfnamefont
  {R.}~\bibnamefont {Clarke}},  \emph {et~al.},\ }\bibfield  {title} {\enquote
  {\bibinfo {title} {Electric field detection in laser-plasma interaction
  experiments via the proton imaging technique},}\ }\href@noop {} {\bibfield
  {journal} {\bibinfo  {journal} {Physics of Plasmas}\ }\textbf {\bibinfo
  {volume} {9}},\ \bibinfo {pages} {2214--2220} (\bibinfo {year}
  {2002})}\BibitemShut {NoStop}%
\bibitem [{\citenamefont {Borghesi}\ \emph {et~al.}(2003)\citenamefont
  {Borghesi}, \citenamefont {Romagnani}, \citenamefont {Schiavi}, \citenamefont
  {Campbell}, \citenamefont {Haines}, \citenamefont {Willi}, \citenamefont
  {Mackinnon}, \citenamefont {Galimberti}, \citenamefont {Gizzi}, \citenamefont
  {Clarke} \emph {et~al.}}]{borghesi2003measurement}%
  \BibitemOpen
  \bibfield  {author} {\bibinfo {author} {\bibfnamefont {M.}~\bibnamefont
  {Borghesi}}, \bibinfo {author} {\bibfnamefont {L.}~\bibnamefont {Romagnani}},
  \bibinfo {author} {\bibfnamefont {A.}~\bibnamefont {Schiavi}}, \bibinfo
  {author} {\bibfnamefont {D.}~\bibnamefont {Campbell}}, \bibinfo {author}
  {\bibfnamefont {M.}~\bibnamefont {Haines}}, \bibinfo {author} {\bibfnamefont
  {O.}~\bibnamefont {Willi}}, \bibinfo {author} {\bibfnamefont
  {A.}~\bibnamefont {Mackinnon}}, \bibinfo {author} {\bibfnamefont
  {M.}~\bibnamefont {Galimberti}}, \bibinfo {author} {\bibfnamefont
  {L.}~\bibnamefont {Gizzi}}, \bibinfo {author} {\bibfnamefont
  {R.}~\bibnamefont {Clarke}},  \emph {et~al.},\ }\bibfield  {title} {\enquote
  {\bibinfo {title} {Measurement of highly transient electrical charging
  following high-intensity laser--solid interaction},}\ }\href@noop {}
  {\bibfield  {journal} {\bibinfo  {journal} {Applied Physics Letters}\
  }\textbf {\bibinfo {volume} {82}},\ \bibinfo {pages} {1529--1531} (\bibinfo
  {year} {2003})}\BibitemShut {NoStop}%
\bibitem [{\citenamefont {Wilks}\ \emph {et~al.}(2001)\citenamefont {Wilks},
  \citenamefont {Langdon}, \citenamefont {Cowan}, \citenamefont {Roth},
  \citenamefont {Singh}, \citenamefont {Hatchett}, \citenamefont {Key},
  \citenamefont {Pennington}, \citenamefont {MacKinnon},\ and\ \citenamefont
  {Snavely}}]{wilks2001energetic}%
  \BibitemOpen
  \bibfield  {author} {\bibinfo {author} {\bibfnamefont {S.}~\bibnamefont
  {Wilks}}, \bibinfo {author} {\bibfnamefont {A.}~\bibnamefont {Langdon}},
  \bibinfo {author} {\bibfnamefont {T.}~\bibnamefont {Cowan}}, \bibinfo
  {author} {\bibfnamefont {M.}~\bibnamefont {Roth}}, \bibinfo {author}
  {\bibfnamefont {M.}~\bibnamefont {Singh}}, \bibinfo {author} {\bibfnamefont
  {S.}~\bibnamefont {Hatchett}}, \bibinfo {author} {\bibfnamefont
  {M.}~\bibnamefont {Key}}, \bibinfo {author} {\bibfnamefont {D.}~\bibnamefont
  {Pennington}}, \bibinfo {author} {\bibfnamefont {A.}~\bibnamefont
  {MacKinnon}}, \ and\ \bibinfo {author} {\bibfnamefont {R.}~\bibnamefont
  {Snavely}},\ }\bibfield  {title} {\enquote {\bibinfo {title} {Energetic
  proton generation in ultra-intense laser--solid interactions},}\ }\href@noop
  {} {\bibfield  {journal} {\bibinfo  {journal} {Physics of plasmas}\ }\textbf
  {\bibinfo {volume} {8}},\ \bibinfo {pages} {542--549} (\bibinfo {year}
  {2001})}\BibitemShut {NoStop}%
\bibitem [{\citenamefont {Snavely}\ \emph {et~al.}(2000)\citenamefont
  {Snavely}, \citenamefont {Key}, \citenamefont {Hatchett}, \citenamefont
  {Cowan}, \citenamefont {Roth}, \citenamefont {Phillips}, \citenamefont
  {Stoyer}, \citenamefont {Henry}, \citenamefont {Sangster}, \citenamefont
  {Singh} \emph {et~al.}}]{snavely2000intense}%
  \BibitemOpen
  \bibfield  {author} {\bibinfo {author} {\bibfnamefont {R.}~\bibnamefont
  {Snavely}}, \bibinfo {author} {\bibfnamefont {M.}~\bibnamefont {Key}},
  \bibinfo {author} {\bibfnamefont {S.}~\bibnamefont {Hatchett}}, \bibinfo
  {author} {\bibfnamefont {T.}~\bibnamefont {Cowan}}, \bibinfo {author}
  {\bibfnamefont {M.}~\bibnamefont {Roth}}, \bibinfo {author} {\bibfnamefont
  {T.}~\bibnamefont {Phillips}}, \bibinfo {author} {\bibfnamefont
  {M.}~\bibnamefont {Stoyer}}, \bibinfo {author} {\bibfnamefont
  {E.}~\bibnamefont {Henry}}, \bibinfo {author} {\bibfnamefont
  {T.}~\bibnamefont {Sangster}}, \bibinfo {author} {\bibfnamefont
  {M.}~\bibnamefont {Singh}},  \emph {et~al.},\ }\bibfield  {title} {\enquote
  {\bibinfo {title} {Intense high-energy proton beams from petawatt-laser
  irradiation of solids},}\ }\href@noop {} {\bibfield  {journal} {\bibinfo
  {journal} {Physical review letters}\ }\textbf {\bibinfo {volume} {85}},\
  \bibinfo {pages} {2945} (\bibinfo {year} {2000})}\BibitemShut {NoStop}%
\bibitem [{\citenamefont {Mora}(2003)}]{mora2003plasma}%
  \BibitemOpen
  \bibfield  {author} {\bibinfo {author} {\bibfnamefont {P.}~\bibnamefont
  {Mora}},\ }\bibfield  {title} {\enquote {\bibinfo {title} {Plasma expansion
  into a vacuum},}\ }\href@noop {} {\bibfield  {journal} {\bibinfo  {journal}
  {Physical Review Letters}\ }\textbf {\bibinfo {volume} {90}},\ \bibinfo
  {pages} {185002} (\bibinfo {year} {2003})}\BibitemShut {NoStop}%
\bibitem [{\citenamefont {Goswami}\ \emph {et~al.}(2021)\citenamefont
  {Goswami}, \citenamefont {Maity}, \citenamefont {Mandal}, \citenamefont
  {Vashistha},\ and\ \citenamefont {Das}}]{goswami2021ponderomotive}%
  \BibitemOpen
  \bibfield  {author} {\bibinfo {author} {\bibfnamefont {L.~P.}\ \bibnamefont
  {Goswami}}, \bibinfo {author} {\bibfnamefont {S.}~\bibnamefont {Maity}},
  \bibinfo {author} {\bibfnamefont {D.}~\bibnamefont {Mandal}}, \bibinfo
  {author} {\bibfnamefont {A.}~\bibnamefont {Vashistha}}, \ and\ \bibinfo
  {author} {\bibfnamefont {A.}~\bibnamefont {Das}},\ }\bibfield  {title}
  {\enquote {\bibinfo {title} {Ponderomotive force driven mechanism for
  electrostatic wave excitation and energy absorption of electromagnetic waves
  in overdense magnetized plasma},}\ }\href@noop {} {\bibfield  {journal}
  {\bibinfo  {journal} {Plasma Physics and Controlled Fusion}\ }\textbf
  {\bibinfo {volume} {63}},\ \bibinfo {pages} {115003} (\bibinfo {year}
  {2021})}\BibitemShut {NoStop}%
\bibitem [{\citenamefont {Ferri}, \citenamefont {Siminos},\ and\ \citenamefont
  {F{\"u}l{\"o}p}(2019)}]{ferri2019enhanced}%
  \BibitemOpen
  \bibfield  {author} {\bibinfo {author} {\bibfnamefont {J.}~\bibnamefont
  {Ferri}}, \bibinfo {author} {\bibfnamefont {E.}~\bibnamefont {Siminos}}, \
  and\ \bibinfo {author} {\bibfnamefont {T.}~\bibnamefont {F{\"u}l{\"o}p}},\
  }\bibfield  {title} {\enquote {\bibinfo {title} {Enhanced target normal
  sheath acceleration using colliding laser pulses},}\ }\href@noop {}
  {\bibfield  {journal} {\bibinfo  {journal} {Communications Physics}\ }\textbf
  {\bibinfo {volume} {2}},\ \bibinfo {pages} {1--8} (\bibinfo {year}
  {2019})}\BibitemShut {NoStop}%
\bibitem [{\citenamefont {Klimo}\ \emph {et~al.}(2011)\citenamefont {Klimo},
  \citenamefont {Psikal}, \citenamefont {Limpouch}, \citenamefont {Proska},
  \citenamefont {Novotny}, \citenamefont {Ceccotti}, \citenamefont {Floquet},\
  and\ \citenamefont {Kawata}}]{klimo2011short}%
  \BibitemOpen
  \bibfield  {author} {\bibinfo {author} {\bibfnamefont {O.}~\bibnamefont
  {Klimo}}, \bibinfo {author} {\bibfnamefont {J.}~\bibnamefont {Psikal}},
  \bibinfo {author} {\bibfnamefont {J.}~\bibnamefont {Limpouch}}, \bibinfo
  {author} {\bibfnamefont {J.}~\bibnamefont {Proska}}, \bibinfo {author}
  {\bibfnamefont {F.}~\bibnamefont {Novotny}}, \bibinfo {author} {\bibfnamefont
  {T.}~\bibnamefont {Ceccotti}}, \bibinfo {author} {\bibfnamefont
  {V.}~\bibnamefont {Floquet}}, \ and\ \bibinfo {author} {\bibfnamefont
  {S.}~\bibnamefont {Kawata}},\ }\bibfield  {title} {\enquote {\bibinfo {title}
  {Short pulse laser interaction with micro-structured targets: simulations of
  laser absorption and ion acceleration},}\ }\href@noop {} {\bibfield
  {journal} {\bibinfo  {journal} {New journal of physics}\ }\textbf {\bibinfo
  {volume} {13}},\ \bibinfo {pages} {053028} (\bibinfo {year}
  {2011})}\BibitemShut {NoStop}%
\bibitem [{\citenamefont {Zou}\ \emph {et~al.}(2019)\citenamefont {Zou},
  \citenamefont {Yu}, \citenamefont {Jiang}, \citenamefont {Yu}, \citenamefont
  {Chen}, \citenamefont {Deng}, \citenamefont {Yu}, \citenamefont {Yin},
  \citenamefont {Shao}, \citenamefont {Zhuo} \emph
  {et~al.}}]{zou2019enhancement}%
  \BibitemOpen
  \bibfield  {author} {\bibinfo {author} {\bibfnamefont {D.}~\bibnamefont
  {Zou}}, \bibinfo {author} {\bibfnamefont {D.}~\bibnamefont {Yu}}, \bibinfo
  {author} {\bibfnamefont {X.}~\bibnamefont {Jiang}}, \bibinfo {author}
  {\bibfnamefont {M.}~\bibnamefont {Yu}}, \bibinfo {author} {\bibfnamefont
  {Z.}~\bibnamefont {Chen}}, \bibinfo {author} {\bibfnamefont {Z.}~\bibnamefont
  {Deng}}, \bibinfo {author} {\bibfnamefont {T.}~\bibnamefont {Yu}}, \bibinfo
  {author} {\bibfnamefont {Y.}~\bibnamefont {Yin}}, \bibinfo {author}
  {\bibfnamefont {F.}~\bibnamefont {Shao}}, \bibinfo {author} {\bibfnamefont
  {H.}~\bibnamefont {Zhuo}},  \emph {et~al.},\ }\bibfield  {title} {\enquote
  {\bibinfo {title} {Enhancement of target normal sheath acceleration in laser
  multi-channel target interaction},}\ }\href@noop {} {\bibfield  {journal}
  {\bibinfo  {journal} {Physics of Plasmas}\ }\textbf {\bibinfo {volume}
  {26}},\ \bibinfo {pages} {123105} (\bibinfo {year} {2019})}\BibitemShut
  {NoStop}%
\bibitem [{\citenamefont {Andreev}\ \emph {et~al.}(2011)\citenamefont
  {Andreev}, \citenamefont {Kumar}, \citenamefont {Platonov},\ and\
  \citenamefont {Pukhov}}]{andreev2011efficient}%
  \BibitemOpen
  \bibfield  {author} {\bibinfo {author} {\bibfnamefont {A.}~\bibnamefont
  {Andreev}}, \bibinfo {author} {\bibfnamefont {N.}~\bibnamefont {Kumar}},
  \bibinfo {author} {\bibfnamefont {K.}~\bibnamefont {Platonov}}, \ and\
  \bibinfo {author} {\bibfnamefont {A.}~\bibnamefont {Pukhov}},\ }\bibfield
  {title} {\enquote {\bibinfo {title} {Efficient generation of fast ions from
  surface modulated nanostructure targets irradiated by high intensity
  short-pulse lasers},}\ }\href@noop {} {\bibfield  {journal} {\bibinfo
  {journal} {Physics of Plasmas}\ }\textbf {\bibinfo {volume} {18}},\ \bibinfo
  {pages} {103103} (\bibinfo {year} {2011})}\BibitemShut {NoStop}%
\bibitem [{\citenamefont {Khan}\ and\ \citenamefont
  {Saxena}(2023)}]{khan2023enhanced}%
  \BibitemOpen
  \bibfield  {author} {\bibinfo {author} {\bibfnamefont {I.}~\bibnamefont
  {Khan}}\ and\ \bibinfo {author} {\bibfnamefont {V.}~\bibnamefont {Saxena}},\
  }\bibfield  {title} {\enquote {\bibinfo {title} {Enhanced target normal
  sheath acceleration with a grooved hydrocarbon target},}\ }\href@noop {}
  {\bibfield  {journal} {\bibinfo  {journal} {Physics of Plasmas}\ }\textbf
  {\bibinfo {volume} {30}} (\bibinfo {year} {2023})}\BibitemShut {NoStop}%
\bibitem [{\citenamefont {Ferri}\ \emph {et~al.}(2020)\citenamefont {Ferri},
  \citenamefont {Siminos}, \citenamefont {Gremillet},\ and\ \citenamefont
  {F{\"u}l{\"o}p}}]{ferri2020effects}%
  \BibitemOpen
  \bibfield  {author} {\bibinfo {author} {\bibfnamefont {J.}~\bibnamefont
  {Ferri}}, \bibinfo {author} {\bibfnamefont {E.}~\bibnamefont {Siminos}},
  \bibinfo {author} {\bibfnamefont {L.}~\bibnamefont {Gremillet}}, \ and\
  \bibinfo {author} {\bibfnamefont {T.}~\bibnamefont {F{\"u}l{\"o}p}},\
  }\bibfield  {title} {\enquote {\bibinfo {title} {Effects of oblique incidence
  and colliding pulses on laser-driven proton acceleration from
  relativistically transparent ultrathin targets},}\ }\href@noop {} {\bibfield
  {journal} {\bibinfo  {journal} {Journal of Plasma Physics}\ }\textbf
  {\bibinfo {volume} {86}},\ \bibinfo {pages} {905860505} (\bibinfo {year}
  {2020})}\BibitemShut {NoStop}%
\bibitem [{\citenamefont {Brunel}(1987)}]{brunel1987not}%
  \BibitemOpen
  \bibfield  {author} {\bibinfo {author} {\bibfnamefont {F.}~\bibnamefont
  {Brunel}},\ }\bibfield  {title} {\enquote {\bibinfo {title} {Not-so-resonant,
  resonant absorption},}\ }\href@noop {} {\bibfield  {journal} {\bibinfo
  {journal} {Physical review letters}\ }\textbf {\bibinfo {volume} {59}},\
  \bibinfo {pages} {52} (\bibinfo {year} {1987})}\BibitemShut {NoStop}%
\bibitem [{\citenamefont {Ferri}\ \emph {et~al.}(2018)\citenamefont {Ferri},
  \citenamefont {Senje}, \citenamefont {Dalui}, \citenamefont {Svensson},
  \citenamefont {Aurand}, \citenamefont {Hansson}, \citenamefont {Persson},
  \citenamefont {Lundh}, \citenamefont {Wahlstr{\"o}m}, \citenamefont
  {Gremillet} \emph {et~al.}}]{ferri2018proton}%
  \BibitemOpen
  \bibfield  {author} {\bibinfo {author} {\bibfnamefont {J.}~\bibnamefont
  {Ferri}}, \bibinfo {author} {\bibfnamefont {L.}~\bibnamefont {Senje}},
  \bibinfo {author} {\bibfnamefont {M.}~\bibnamefont {Dalui}}, \bibinfo
  {author} {\bibfnamefont {K.}~\bibnamefont {Svensson}}, \bibinfo {author}
  {\bibfnamefont {B.}~\bibnamefont {Aurand}}, \bibinfo {author} {\bibfnamefont
  {M.}~\bibnamefont {Hansson}}, \bibinfo {author} {\bibfnamefont
  {A.}~\bibnamefont {Persson}}, \bibinfo {author} {\bibfnamefont
  {O.}~\bibnamefont {Lundh}}, \bibinfo {author} {\bibfnamefont {C.-G.}\
  \bibnamefont {Wahlstr{\"o}m}}, \bibinfo {author} {\bibfnamefont
  {L.}~\bibnamefont {Gremillet}},  \emph {et~al.},\ }\bibfield  {title}
  {\enquote {\bibinfo {title} {Proton acceleration by a pair of successive
  ultraintense femtosecond laser pulses},}\ }\href@noop {} {\bibfield
  {journal} {\bibinfo  {journal} {Physics of Plasmas}\ }\textbf {\bibinfo
  {volume} {25}},\ \bibinfo {pages} {043115} (\bibinfo {year}
  {2018})}\BibitemShut {NoStop}%
\bibitem [{\citenamefont {Scott}\ \emph {et~al.}(2012)\citenamefont {Scott},
  \citenamefont {Green}, \citenamefont {Bagnoud}, \citenamefont {Brabetz},
  \citenamefont {Brenner}, \citenamefont {Carroll}, \citenamefont {MacLellan},
  \citenamefont {Robinson}, \citenamefont {Roth}, \citenamefont {Spindloe}
  \emph {et~al.}}]{scott2012multi}%
  \BibitemOpen
  \bibfield  {author} {\bibinfo {author} {\bibfnamefont {G.}~\bibnamefont
  {Scott}}, \bibinfo {author} {\bibfnamefont {J.}~\bibnamefont {Green}},
  \bibinfo {author} {\bibfnamefont {V.}~\bibnamefont {Bagnoud}}, \bibinfo
  {author} {\bibfnamefont {C.}~\bibnamefont {Brabetz}}, \bibinfo {author}
  {\bibfnamefont {C.}~\bibnamefont {Brenner}}, \bibinfo {author} {\bibfnamefont
  {D.}~\bibnamefont {Carroll}}, \bibinfo {author} {\bibfnamefont
  {D.}~\bibnamefont {MacLellan}}, \bibinfo {author} {\bibfnamefont
  {A.}~\bibnamefont {Robinson}}, \bibinfo {author} {\bibfnamefont
  {M.}~\bibnamefont {Roth}}, \bibinfo {author} {\bibfnamefont {C.}~\bibnamefont
  {Spindloe}},  \emph {et~al.},\ }\bibfield  {title} {\enquote {\bibinfo
  {title} {Multi-pulse enhanced laser ion acceleration using plasma half cavity
  targets},}\ }\href@noop {} {\bibfield  {journal} {\bibinfo  {journal}
  {Applied physics letters}\ }\textbf {\bibinfo {volume} {101}},\ \bibinfo
  {pages} {024101} (\bibinfo {year} {2012})}\BibitemShut {NoStop}%
\bibitem [{\citenamefont {Markey}\ \emph {et~al.}(2010)\citenamefont {Markey},
  \citenamefont {McKenna}, \citenamefont {Brenner}, \citenamefont {Carroll},
  \citenamefont {G{\"u}nther}, \citenamefont {Harres}, \citenamefont {Kar},
  \citenamefont {Lancaster}, \citenamefont {N{\"u}rnberg}, \citenamefont
  {Quinn} \emph {et~al.}}]{markey2010spectral}%
  \BibitemOpen
  \bibfield  {author} {\bibinfo {author} {\bibfnamefont {K.}~\bibnamefont
  {Markey}}, \bibinfo {author} {\bibfnamefont {P.}~\bibnamefont {McKenna}},
  \bibinfo {author} {\bibfnamefont {C.}~\bibnamefont {Brenner}}, \bibinfo
  {author} {\bibfnamefont {D.}~\bibnamefont {Carroll}}, \bibinfo {author}
  {\bibfnamefont {M.}~\bibnamefont {G{\"u}nther}}, \bibinfo {author}
  {\bibfnamefont {K.}~\bibnamefont {Harres}}, \bibinfo {author} {\bibfnamefont
  {S.}~\bibnamefont {Kar}}, \bibinfo {author} {\bibfnamefont {K.}~\bibnamefont
  {Lancaster}}, \bibinfo {author} {\bibfnamefont {F.}~\bibnamefont
  {N{\"u}rnberg}}, \bibinfo {author} {\bibfnamefont {M.}~\bibnamefont {Quinn}},
   \emph {et~al.},\ }\bibfield  {title} {\enquote {\bibinfo {title} {Spectral
  enhancement in the double pulse regime of laser proton acceleration},}\
  }\href@noop {} {\bibfield  {journal} {\bibinfo  {journal} {Physical review
  letters}\ }\textbf {\bibinfo {volume} {105}},\ \bibinfo {pages} {195008}
  (\bibinfo {year} {2010})}\BibitemShut {NoStop}%
\bibitem [{\citenamefont {Arber}\ \emph {et~al.}(2015)\citenamefont {Arber},
  \citenamefont {Bennett}, \citenamefont {Brady}, \citenamefont
  {Lawrence-Douglas}, \citenamefont {Ramsay}, \citenamefont {Sircombe},
  \citenamefont {Gillies}, \citenamefont {Evans}, \citenamefont {Schmitz},
  \citenamefont {Bell} \emph {et~al.}}]{arber2015contemporary}%
  \BibitemOpen
  \bibfield  {author} {\bibinfo {author} {\bibfnamefont {T.}~\bibnamefont
  {Arber}}, \bibinfo {author} {\bibfnamefont {K.}~\bibnamefont {Bennett}},
  \bibinfo {author} {\bibfnamefont {C.}~\bibnamefont {Brady}}, \bibinfo
  {author} {\bibfnamefont {A.}~\bibnamefont {Lawrence-Douglas}}, \bibinfo
  {author} {\bibfnamefont {M.}~\bibnamefont {Ramsay}}, \bibinfo {author}
  {\bibfnamefont {N.}~\bibnamefont {Sircombe}}, \bibinfo {author}
  {\bibfnamefont {P.}~\bibnamefont {Gillies}}, \bibinfo {author} {\bibfnamefont
  {R.}~\bibnamefont {Evans}}, \bibinfo {author} {\bibfnamefont
  {H.}~\bibnamefont {Schmitz}}, \bibinfo {author} {\bibfnamefont
  {A.}~\bibnamefont {Bell}},  \emph {et~al.},\ }\bibfield  {title} {\enquote
  {\bibinfo {title} {Contemporary particle-in-cell approach to laser-plasma
  modelling},}\ }\href@noop {} {\bibfield  {journal} {\bibinfo  {journal}
  {Plasma Physics and Controlled Fusion}\ }\textbf {\bibinfo {volume} {57}},\
  \bibinfo {pages} {113001} (\bibinfo {year} {2015})}\BibitemShut {NoStop}%
\bibitem [{\citenamefont {Sgattoni}\ \emph {et~al.}(2012)\citenamefont
  {Sgattoni}, \citenamefont {Londrillo}, \citenamefont {Macchi},\ and\
  \citenamefont {Passoni}}]{sgattoni2012laser}%
  \BibitemOpen
  \bibfield  {author} {\bibinfo {author} {\bibfnamefont {A.}~\bibnamefont
  {Sgattoni}}, \bibinfo {author} {\bibfnamefont {P.}~\bibnamefont {Londrillo}},
  \bibinfo {author} {\bibfnamefont {A.}~\bibnamefont {Macchi}}, \ and\ \bibinfo
  {author} {\bibfnamefont {M.}~\bibnamefont {Passoni}},\ }\bibfield  {title}
  {\enquote {\bibinfo {title} {Laser ion acceleration using a solid target
  coupled with a low-density layer},}\ }\href@noop {} {\bibfield  {journal}
  {\bibinfo  {journal} {Physical Review E}\ }\textbf {\bibinfo {volume} {85}},\
  \bibinfo {pages} {036405} (\bibinfo {year} {2012})}\BibitemShut {NoStop}%
\bibitem [{\citenamefont {d'Humi{\`e}res}\ \emph {et~al.}(2013)\citenamefont
  {d'Humi{\`e}res}, \citenamefont {Brantov}, \citenamefont {Yu.~Bychenkov},\
  and\ \citenamefont {Tikhonchuk}}]{d2013optimization}%
  \BibitemOpen
  \bibfield  {author} {\bibinfo {author} {\bibfnamefont {E.}~\bibnamefont
  {d'Humi{\`e}res}}, \bibinfo {author} {\bibfnamefont {A.}~\bibnamefont
  {Brantov}}, \bibinfo {author} {\bibfnamefont {V.}~\bibnamefont
  {Yu.~Bychenkov}}, \ and\ \bibinfo {author} {\bibfnamefont {V.}~\bibnamefont
  {Tikhonchuk}},\ }\bibfield  {title} {\enquote {\bibinfo {title} {Optimization
  of laser-target interaction for proton acceleration},}\ }\href@noop {}
  {\bibfield  {journal} {\bibinfo  {journal} {Physics of Plasmas}\ }\textbf
  {\bibinfo {volume} {20}},\ \bibinfo {pages} {023103} (\bibinfo {year}
  {2013})}\BibitemShut {NoStop}%
\bibitem [{\citenamefont {Stark}\ \emph {et~al.}(2017)\citenamefont {Stark},
  \citenamefont {Yin}, \citenamefont {Albright},\ and\ \citenamefont
  {Guo}}]{stark2017effects}%
  \BibitemOpen
  \bibfield  {author} {\bibinfo {author} {\bibfnamefont {D.~J.}\ \bibnamefont
  {Stark}}, \bibinfo {author} {\bibfnamefont {L.}~\bibnamefont {Yin}}, \bibinfo
  {author} {\bibfnamefont {B.~J.}\ \bibnamefont {Albright}}, \ and\ \bibinfo
  {author} {\bibfnamefont {F.}~\bibnamefont {Guo}},\ }\bibfield  {title}
  {\enquote {\bibinfo {title} {Effects of dimensionality on kinetic simulations
  of laser-ion acceleration in the transparency regime},}\ }\href@noop {}
  {\bibfield  {journal} {\bibinfo  {journal} {Physics of Plasmas}\ }\textbf
  {\bibinfo {volume} {24}},\ \bibinfo {pages} {053103} (\bibinfo {year}
  {2017})}\BibitemShut {NoStop}%
\bibitem [{\citenamefont {Scullion}\ \emph {et~al.}(2017)\citenamefont
  {Scullion}, \citenamefont {Doria}, \citenamefont {Romagnani}, \citenamefont
  {Sgattoni}, \citenamefont {Naughton}, \citenamefont {Symes}, \citenamefont
  {McKenna}, \citenamefont {Macchi}, \citenamefont {Zepf}, \citenamefont {Kar}
  \emph {et~al.}}]{scullion2017polarization}%
  \BibitemOpen
  \bibfield  {author} {\bibinfo {author} {\bibfnamefont {C.}~\bibnamefont
  {Scullion}}, \bibinfo {author} {\bibfnamefont {D.}~\bibnamefont {Doria}},
  \bibinfo {author} {\bibfnamefont {L.}~\bibnamefont {Romagnani}}, \bibinfo
  {author} {\bibfnamefont {A.}~\bibnamefont {Sgattoni}}, \bibinfo {author}
  {\bibfnamefont {K.}~\bibnamefont {Naughton}}, \bibinfo {author}
  {\bibfnamefont {D.}~\bibnamefont {Symes}}, \bibinfo {author} {\bibfnamefont
  {P.}~\bibnamefont {McKenna}}, \bibinfo {author} {\bibfnamefont
  {A.}~\bibnamefont {Macchi}}, \bibinfo {author} {\bibfnamefont
  {M.}~\bibnamefont {Zepf}}, \bibinfo {author} {\bibfnamefont {S.}~\bibnamefont
  {Kar}},  \emph {et~al.},\ }\bibfield  {title} {\enquote {\bibinfo {title}
  {Polarization dependence of bulk ion acceleration from ultrathin foils
  irradiated by high-intensity ultrashort laser pulses},}\ }\href@noop {}
  {\bibfield  {journal} {\bibinfo  {journal} {Physical review letters}\
  }\textbf {\bibinfo {volume} {119}},\ \bibinfo {pages} {054801} (\bibinfo
  {year} {2017})}\BibitemShut {NoStop}%
\bibitem [{\citenamefont {Yang}\ \emph {et~al.}(2016)\citenamefont {Yang},
  \citenamefont {Deng}, \citenamefont {Yu},\ and\ \citenamefont
  {Wang}}]{yang2016high}%
  \BibitemOpen
  \bibfield  {author} {\bibinfo {author} {\bibfnamefont {L.}~\bibnamefont
  {Yang}}, \bibinfo {author} {\bibfnamefont {Z.}~\bibnamefont {Deng}}, \bibinfo
  {author} {\bibfnamefont {M.}~\bibnamefont {Yu}}, \ and\ \bibinfo {author}
  {\bibfnamefont {X.}~\bibnamefont {Wang}},\ }\bibfield  {title} {\enquote
  {\bibinfo {title} {High-charge energetic ions generated by intersecting laser
  pulses},}\ }\href@noop {} {\bibfield  {journal} {\bibinfo  {journal} {Physics
  of Plasmas}\ }\textbf {\bibinfo {volume} {23}},\ \bibinfo {pages} {083106}
  (\bibinfo {year} {2016})}\BibitemShut {NoStop}%
\bibitem [{\citenamefont {Yogo}\ \emph {et~al.}(2015)\citenamefont {Yogo},
  \citenamefont {Bulanov}, \citenamefont {Mori}, \citenamefont {Ogura},
  \citenamefont {Esirkepov}, \citenamefont {Pirozhkov}, \citenamefont
  {Kanasaki}, \citenamefont {Sakaki}, \citenamefont {Fukuda}, \citenamefont
  {Bolton} \emph {et~al.}}]{yogo2015ion}%
  \BibitemOpen
  \bibfield  {author} {\bibinfo {author} {\bibfnamefont {A.}~\bibnamefont
  {Yogo}}, \bibinfo {author} {\bibfnamefont {S.}~\bibnamefont {Bulanov}},
  \bibinfo {author} {\bibfnamefont {M.}~\bibnamefont {Mori}}, \bibinfo {author}
  {\bibfnamefont {K.}~\bibnamefont {Ogura}}, \bibinfo {author} {\bibfnamefont
  {T.~Z.}\ \bibnamefont {Esirkepov}}, \bibinfo {author} {\bibfnamefont
  {A.}~\bibnamefont {Pirozhkov}}, \bibinfo {author} {\bibfnamefont
  {M.}~\bibnamefont {Kanasaki}}, \bibinfo {author} {\bibfnamefont
  {H.}~\bibnamefont {Sakaki}}, \bibinfo {author} {\bibfnamefont
  {Y.}~\bibnamefont {Fukuda}}, \bibinfo {author} {\bibfnamefont
  {P.}~\bibnamefont {Bolton}},  \emph {et~al.},\ }\bibfield  {title} {\enquote
  {\bibinfo {title} {Ion acceleration via ‘nonlinear vacuum heating’by the
  laser pulse obliquely incident on a thin foil target},}\ }\href@noop {}
  {\bibfield  {journal} {\bibinfo  {journal} {Plasma Physics and Controlled
  Fusion}\ }\textbf {\bibinfo {volume} {58}},\ \bibinfo {pages} {025003}
  (\bibinfo {year} {2015})}\BibitemShut {NoStop}%
\bibitem [{\citenamefont {Ceccotti}\ \emph {et~al.}(2013)\citenamefont
  {Ceccotti}, \citenamefont {Floquet}, \citenamefont {Sgattoni}, \citenamefont
  {Bigongiari}, \citenamefont {Klimo}, \citenamefont {Raynaud}, \citenamefont
  {Riconda}, \citenamefont {Heron}, \citenamefont {Baffigi}, \citenamefont
  {Labate} \emph {et~al.}}]{ceccotti2013evidence}%
  \BibitemOpen
  \bibfield  {author} {\bibinfo {author} {\bibfnamefont {T.}~\bibnamefont
  {Ceccotti}}, \bibinfo {author} {\bibfnamefont {V.}~\bibnamefont {Floquet}},
  \bibinfo {author} {\bibfnamefont {A.}~\bibnamefont {Sgattoni}}, \bibinfo
  {author} {\bibfnamefont {A.}~\bibnamefont {Bigongiari}}, \bibinfo {author}
  {\bibfnamefont {O.}~\bibnamefont {Klimo}}, \bibinfo {author} {\bibfnamefont
  {M.}~\bibnamefont {Raynaud}}, \bibinfo {author} {\bibfnamefont
  {C.}~\bibnamefont {Riconda}}, \bibinfo {author} {\bibfnamefont
  {A.}~\bibnamefont {Heron}}, \bibinfo {author} {\bibfnamefont
  {F.}~\bibnamefont {Baffigi}}, \bibinfo {author} {\bibfnamefont
  {L.}~\bibnamefont {Labate}},  \emph {et~al.},\ }\bibfield  {title} {\enquote
  {\bibinfo {title} {Evidence of resonant surface-wave excitation in the
  relativistic regime through measurements of proton acceleration from grating
  targets},}\ }\href@noop {} {\bibfield  {journal} {\bibinfo  {journal}
  {Physical review letters}\ }\textbf {\bibinfo {volume} {111}},\ \bibinfo
  {pages} {185001} (\bibinfo {year} {2013})}\BibitemShut {NoStop}%
\bibitem [{\citenamefont {Floquet}\ \emph {et~al.}(2013)\citenamefont
  {Floquet}, \citenamefont {Klimo}, \citenamefont {Psikal}, \citenamefont
  {Velyhan}, \citenamefont {Limpouch}, \citenamefont {Proska}, \citenamefont
  {Novotny}, \citenamefont {Stolcova}, \citenamefont {Macchi}, \citenamefont
  {Sgattoni} \emph {et~al.}}]{floquet2013micro}%
  \BibitemOpen
  \bibfield  {author} {\bibinfo {author} {\bibfnamefont {V.}~\bibnamefont
  {Floquet}}, \bibinfo {author} {\bibfnamefont {O.}~\bibnamefont {Klimo}},
  \bibinfo {author} {\bibfnamefont {J.}~\bibnamefont {Psikal}}, \bibinfo
  {author} {\bibfnamefont {A.}~\bibnamefont {Velyhan}}, \bibinfo {author}
  {\bibfnamefont {J.}~\bibnamefont {Limpouch}}, \bibinfo {author}
  {\bibfnamefont {J.}~\bibnamefont {Proska}}, \bibinfo {author} {\bibfnamefont
  {F.}~\bibnamefont {Novotny}}, \bibinfo {author} {\bibfnamefont
  {L.}~\bibnamefont {Stolcova}}, \bibinfo {author} {\bibfnamefont
  {A.}~\bibnamefont {Macchi}}, \bibinfo {author} {\bibfnamefont
  {A.}~\bibnamefont {Sgattoni}},  \emph {et~al.},\ }\bibfield  {title}
  {\enquote {\bibinfo {title} {Micro-sphere layered targets efficiency in laser
  driven proton acceleration},}\ }\href@noop {} {\bibfield  {journal} {\bibinfo
   {journal} {Journal of Applied Physics}\ }\textbf {\bibinfo {volume} {114}},\
  \bibinfo {pages} {083305} (\bibinfo {year} {2013})}\BibitemShut {NoStop}%
\bibitem [{\citenamefont {Rahman}\ \emph {et~al.}(2021)\citenamefont {Rahman},
  \citenamefont {Smith}, \citenamefont {Ngirmang},\ and\ \citenamefont
  {Orban}}]{rahman2021particle}%
  \BibitemOpen
  \bibfield  {author} {\bibinfo {author} {\bibfnamefont {N.}~\bibnamefont
  {Rahman}}, \bibinfo {author} {\bibfnamefont {J.~R.}\ \bibnamefont {Smith}},
  \bibinfo {author} {\bibfnamefont {G.~K.}\ \bibnamefont {Ngirmang}}, \ and\
  \bibinfo {author} {\bibfnamefont {C.}~\bibnamefont {Orban}},\ }\bibfield
  {title} {\enquote {\bibinfo {title} {Particle-in-cell modeling of a potential
  demonstration experiment for double pulse enhanced target normal sheath
  acceleration},}\ }\href@noop {} {\bibfield  {journal} {\bibinfo  {journal}
  {Physics of Plasmas}\ }\textbf {\bibinfo {volume} {28}},\ \bibinfo {pages}
  {073103} (\bibinfo {year} {2021})}\BibitemShut {NoStop}%
\bibitem [{\citenamefont {Yao}\ \emph {et~al.}(2022)\citenamefont {Yao},
  \citenamefont {Nakatsutsumi}, \citenamefont {Buffechoux}, \citenamefont
  {Antici}, \citenamefont {Borghesi}, \citenamefont {Chen}, \citenamefont
  {d'Humi{\`e}res}, \citenamefont {Gremillet}, \citenamefont {Heathcote},
  \citenamefont {Horn{\`y}} \emph {et~al.}}]{yao2022optimizing}%
  \BibitemOpen
  \bibfield  {author} {\bibinfo {author} {\bibfnamefont {W.}~\bibnamefont
  {Yao}}, \bibinfo {author} {\bibfnamefont {M.}~\bibnamefont {Nakatsutsumi}},
  \bibinfo {author} {\bibfnamefont {S.}~\bibnamefont {Buffechoux}}, \bibinfo
  {author} {\bibfnamefont {P.}~\bibnamefont {Antici}}, \bibinfo {author}
  {\bibfnamefont {M.}~\bibnamefont {Borghesi}}, \bibinfo {author}
  {\bibfnamefont {S.~N.}\ \bibnamefont {Chen}}, \bibinfo {author}
  {\bibfnamefont {E.}~\bibnamefont {d'Humi{\`e}res}}, \bibinfo {author}
  {\bibfnamefont {L.}~\bibnamefont {Gremillet}}, \bibinfo {author}
  {\bibfnamefont {R.}~\bibnamefont {Heathcote}}, \bibinfo {author}
  {\bibfnamefont {V.}~\bibnamefont {Horn{\`y}}},  \emph {et~al.},\ }\bibfield
  {title} {\enquote {\bibinfo {title} {Optimizing the laser coupling, matter
  heating, and particle acceleration from solids by using multiplexed
  ultraintense lasers},}\ }\href@noop {} {\bibfield  {journal} {\bibinfo
  {journal} {arXiv preprint arXiv:2208.06272}\ } (\bibinfo {year}
  {2022})}\BibitemShut {NoStop}%
\bibitem [{\citenamefont {Ji}\ \emph {et~al.}(2016)\citenamefont {Ji},
  \citenamefont {Snyder}, \citenamefont {Pukhov}, \citenamefont {Freeman},\
  and\ \citenamefont {Akli}}]{ji2016towards}%
  \BibitemOpen
  \bibfield  {author} {\bibinfo {author} {\bibfnamefont {L.}~\bibnamefont
  {Ji}}, \bibinfo {author} {\bibfnamefont {J.}~\bibnamefont {Snyder}}, \bibinfo
  {author} {\bibfnamefont {A.}~\bibnamefont {Pukhov}}, \bibinfo {author}
  {\bibfnamefont {R.}~\bibnamefont {Freeman}}, \ and\ \bibinfo {author}
  {\bibfnamefont {K.}~\bibnamefont {Akli}},\ }\bibfield  {title} {\enquote
  {\bibinfo {title} {Towards manipulating relativistic laser pulses with
  micro-tube plasma lenses},}\ }\href@noop {} {\bibfield  {journal} {\bibinfo
  {journal} {Scientific reports}\ }\textbf {\bibinfo {volume} {6}},\ \bibinfo
  {pages} {1--7} (\bibinfo {year} {2016})}\BibitemShut {NoStop}%
\bibitem [{\citenamefont {Snyder}\ \emph {et~al.}(2019)\citenamefont {Snyder},
  \citenamefont {Ji}, \citenamefont {George}, \citenamefont {Willis},
  \citenamefont {Cochran}, \citenamefont {Daskalova}, \citenamefont {Handler},
  \citenamefont {Rubin}, \citenamefont {Poole}, \citenamefont {Nasir} \emph
  {et~al.}}]{snyder2019relativistic}%
  \BibitemOpen
  \bibfield  {author} {\bibinfo {author} {\bibfnamefont {J.}~\bibnamefont
  {Snyder}}, \bibinfo {author} {\bibfnamefont {L.}~\bibnamefont {Ji}}, \bibinfo
  {author} {\bibfnamefont {K.~M.}\ \bibnamefont {George}}, \bibinfo {author}
  {\bibfnamefont {C.}~\bibnamefont {Willis}}, \bibinfo {author} {\bibfnamefont
  {G.~E.}\ \bibnamefont {Cochran}}, \bibinfo {author} {\bibfnamefont
  {R.}~\bibnamefont {Daskalova}}, \bibinfo {author} {\bibfnamefont
  {A.}~\bibnamefont {Handler}}, \bibinfo {author} {\bibfnamefont
  {T.}~\bibnamefont {Rubin}}, \bibinfo {author} {\bibfnamefont {P.~L.}\
  \bibnamefont {Poole}}, \bibinfo {author} {\bibfnamefont {D.}~\bibnamefont
  {Nasir}},  \emph {et~al.},\ }\bibfield  {title} {\enquote {\bibinfo {title}
  {Relativistic laser driven electron accelerator using micro-channel plasma
  targets},}\ }\href@noop {} {\bibfield  {journal} {\bibinfo  {journal}
  {Physics of Plasmas}\ }\textbf {\bibinfo {volume} {26}},\ \bibinfo {pages}
  {033110} (\bibinfo {year} {2019})}\BibitemShut {NoStop}%
\bibitem [{\citenamefont {Zhu}\ \emph {et~al.}(2022)\citenamefont {Zhu},
  \citenamefont {Liu}, \citenamefont {Chen}, \citenamefont {Weng},
  \citenamefont {McKenna}, \citenamefont {Sheng},\ and\ \citenamefont
  {Zhang}}]{zhu2022bunched}%
  \BibitemOpen
  \bibfield  {author} {\bibinfo {author} {\bibfnamefont {X.-L.}\ \bibnamefont
  {Zhu}}, \bibinfo {author} {\bibfnamefont {W.-Y.}\ \bibnamefont {Liu}},
  \bibinfo {author} {\bibfnamefont {M.}~\bibnamefont {Chen}}, \bibinfo {author}
  {\bibfnamefont {S.-M.}\ \bibnamefont {Weng}}, \bibinfo {author}
  {\bibfnamefont {P.}~\bibnamefont {McKenna}}, \bibinfo {author} {\bibfnamefont
  {Z.-M.}\ \bibnamefont {Sheng}}, \ and\ \bibinfo {author} {\bibfnamefont
  {J.}~\bibnamefont {Zhang}},\ }\bibfield  {title} {\enquote {\bibinfo {title}
  {Bunched proton acceleration from a laser-irradiated cone target},}\
  }\href@noop {} {\bibfield  {journal} {\bibinfo  {journal} {Physical Review
  Applied}\ }\textbf {\bibinfo {volume} {18}},\ \bibinfo {pages} {044051}
  (\bibinfo {year} {2022})}\BibitemShut {NoStop}%
\end{thebibliography}
%


\end{document}